\documentclass[submission,phys]{lib/SciPost} 

\usepackage{graphicx}
\usepackage{comment}
\usepackage{txfonts}
\usepackage{stfloats}

\newcommand{\MSun}{\mbox{$\mathrm{{M}}_\odot$}}

\newcommand{\MJup}{\mbox{$\mathrm{{M}}_{\rm Jup}$}}

\def\apgt{\ {\raise-.5ex\hbox{$\buildrel>\over\sim$}}\ }
\def\aplt{\ {\raise-.5ex\hbox{$\buildrel<\over\sim$}}\ }
\def\lteq{\ {\raise-.5ex\hbox{$\buildrel<\over-$}}\ }

\newcommand{\jumbo}{\mbox{JuMBO}}
\newcommand{\jumbos}{\mbox{JuMBOs}}
\input{lib/journals.def}

\begin{document} 

\begin{center}{\Large \textbf{
      The origin and evolution of wide Jupiter Mass Binary Objects in young stellar clusters
    }}
\end{center} 

\begin{center}
  Simon F. Portegies Zwart$^{*}$,\\
  Erwan Hochart
\end{center} 

\begin{center}
Leiden Observatory, Leiden University, PO Box 9513, 2300 RA, Leiden, The Netherlands\\
* spz@strw.leidenuniv.nl
\end{center}

\section*{Abstract} {\bf 
      The recently observed population of 540 free-floating
      Jupiter-mass objects, including 40 dynamically soft pairs, and
      two triples, in the Trapezium cluster have raised interesting
      questions on their formation and evolution.  We test various
      scenarios for the origin and survivability of these free
      floating Jupiter-mass objects and Jupiter-mass Binary Objects
      (JuMBOs) in the Trapezium cluster.  The numerical calculations
      are performed by direct $N$-body integration of the stars and
      planets in the Trapezium cluster starting with a wide variety of
      planets in various configurations. We discuss four models:
      $\mathcal{SPP}$, in which selected stars have two outer orbiting
      Jupiter-mass planets; $\mathcal{SPM}$, where selected stars are
      orbited by Jupiter-mass planet-moon pairs; $\mathcal{ISF}$ in
      which \jumbos\, form in situ with the stars, and
      $\mathcal{FFC}$, where we introduce a population of
      free-floating single Jupiter-mass objects, but no initialised
      binaries.  Models $\mathcal{FFC}$ and $\mathcal{SPP}$ fail to
      produce enough \jumbos. Models $\mathcal{SPM}$ can produce
      sufficient free-floaters and \jumbos\/, but requires unusually
      wide orbits for the planet-moon system around the star. The
      observed \jumbos\ and free-floating Jupiter-mass objects in the
      Trapezium cluster are best reproduced if they formed in pairs
      and as free-floaters together with the other stars in a smooth
      (Plummer) density profile with a virial radius of $\sim
      0.5$\,pc.  A fractal (with fractal dimension 1.6) stellar
      density distribution also works, but requires relatively recent
      formations ($\apgt 0.2$\,Myr after the other stars formed) or a
      high ($\apgt 50$\%) initial binary fraction.  This would make
      the primordial binary fraction of \jumbos\, even higher than the
      already large observation fraction of $\sim 8$\,\% (42/540). The
      fraction of \jumbos\, will continue to drop with time, and the
      lack of \jumbos\ in Upper Scorpius could then result in its
      higher age, causing more \jumbos\, to be ionized. We then also
      predict that the interstellar density of Jupiter-mass objects
      (mostly singles with some $\sim 2$\% lucky surviving binaries)
      is $\sim 0.05$\, per pc$^{-3}$ (or around 0.24 per star). }

\section{Introduction}

Recently \cite{2023arXiv231001231P} reported the discovery of 42
Jupiter-Mass Binary Objects (\jumbos) in the direction of the
Trapezium cluster.  Their component masses range between 0.6\,\MJup\,
and 14\,\MJup\, and they have projected separations between 25\,au and
380\,au.  Two of these objects have a nearby tertiary Jupiter-mass
companion.  They also observed a population of 540 single objects in
the same mass range. This discovery initiates
discussions on the origin and survivability of weakly bound
Jupiter-mass pairs in a clustered environment.

Free-floating Jupiter-mass objects have already been detected, with the first 
sightings in the direction of the Trapezium cluster more than twenty years ago \cite{2000Sci...290..103Z,2000MNRAS.314..858L,2000AGM....17..A11M}.
Since then, many more have been found, for example, in the young
clustered environment of Upper Scorpius \cite{2022NatAs...6...89M},
and through gravitational microlensing surveys in the direction of the
Galactic bulge \cite{2011Natur.473..349S}.  Their abundance may be as
high as $1.9^{+1.3}_{-0.8}$ per star \cite{2011Natur.473..349S},
although a considerable fraction of these could be in wide orbits
around a parent star, or have masses $\ll \MJup$.

The origin of these free-floating planets has been debated in
\cite{2023Ap&SS.368...17M}. Generally, star formation via the collapse
of a molecular cloud through gravitational instability leads to
objects considerably more massive than Jupiter
\cite{1976MNRAS.176..367L,2005A&A...430.1059B}, and in disks, planets
tend to form with lower masses. The formation of extremely low-mass
stars, however, remains an active field of research. Theory developed
by \cite{2006A&A...458..817W} introduces the possibility for in situ
formation of objects with Jupiter masses. Even so, the current
consensus remains that the large population of Jupiter-mass
free-floaters arises from fully packed planetary systems
\cite{2023arXiv231015603C}, or as ejectees from their birth planetary
system following encounters with other stars in the cluster
\cite{2019A&A...624A.120V}.  Single Jupiter-mass free-floating objects
then originally formed in a disk around a star to become single later
in time
\cite{1996Sci...274..954R,2015MNRAS.453.2759Z,2002ApJ...565.1251H,
  2017MNRAS.470.4337C, 2019MNRAS.489.2280F,2019A&A...624A.120V}.  The
number of super-Jupiter mass free-floating objects formed in this way
are predicted to be on the order of one ($\sim 0.71$) per star
\cite{2019A&A...624A.120V}, but lower-mass free-floaters orphaned this
way may be much more abundant \cite{2002ApJ...565.1251H}; The origin
of relatively massive free-floaters through dynamical phenomena is
further complicated by the tendency for lower-mass planets to be more
prone to ejections \cite{2001Icar..150..303F,2013MNRAS.433..867H,
  2019MNRAS.489.2280F,2020MNRAS.497.1807S}.

Explaining the observed abundance and mass-function of single
free-floating Jupiter-mass objects is difficult.  In particular, the
large population of objects in Upper Scorpius challenges the formation
channels. The recent discovery of a large population of paired
free-floaters complicates matters even further and puts strong
constraints on their origin. So far, binary free-floating
planetary-mass objects have been rare, and were only discovered in tight
(few au) orbits \cite{2021ApJS..253....7K}, including:  
\begin{itemize}
\item[$\bullet$]2MASS J11193254-1137466 AB: a $5$ to 10\,\MJup\,
  primary in a $a=3.6\pm0.9$\,au orbit \cite{2017ApJ...843L...4B}.
\item[$\bullet$]WISE 1828+2650: a 3 to 6\,\MJup\, primary with a
  5\,\MJup\ companion in an $\apgt 0.5$\,au orbit
  \cite{2013ApJ...764..101B}.
\item[$\bullet$] WISE J0336-014: a $8.5$ to
  $18$\,\MJup\ primary with a $5$ to $11.5$\,\MJup\, companion in a
  $0.9^{+0.05}_{-0.09}$\,au orbit \cite{2023ApJ...947L..30C}.
\item[$\bullet$]2MASS J0013-1143 discovered by \cite{2017AJ....154..112K} and
  suspected to be a binary by \cite{2019A&A...629A.145E}.
\end{itemize}

Such tight pairs could have formed as binary planets (or planet-moon
pairs) orbiting stars, before being dislodged from their parents
\cite{2016ApJ...819..125C}.  If only a few of these objects were
discovered in tight orbits, such an exotic scenario would explain
their existence reasonably well, but the discovery of a rich
population of $42$ wide \jumbos\, \cite{2023arXiv231001231P} requires
a more thorough study of their origin.

Assuming a dynamical history, we perform direct $N$-body simulations of a 
Trapezium-like star cluster with primordial Jupiter-mass objects (JMO) and \jumbos. 
Our simulations focus on four models that could explain the abundance,
and properties (mass and separation distributions) of the observed JMOs in the
cluster. Alternative to forming in situ (scenario ${\cal ISF}$), 
one can naively imagine three mechanisms to
form \jumbos. \cite{2023arXiv231006016W} argued that hierarchical planetary 
systems could explain these binaries. Here, the
outer two planets get stripped by a passing star during a close
encounter. The two ejected planets would lead to a population of free-floating planets, 
but could also explain the observed population of
\jumbos.  We call this scenario ${\cal SPP}$ (for star planet-planet).

Another possibility is that \jumbos\, result from the ejection of planet-moon
pairs (or binary planets) originally orbiting some star.  We call this 
scenario ${\cal SPM}$, for star planet-moon.
Finally, we explore the hypothesis whether a sufficiently large
population of free-floating JMOs could lead to a
population of \jumbos\, by dynamical capture of one JMO by another.
We call this scenario ${\cal FFC}$ (free-floating capture). A similar
scenario was proposed by \cite{2010MNRAS.404.1835K} for explaining
very wide stellar pairs, but the model also works for
wide planetary orbits \cite{2012ApJ...750...83P,2018MNRAS.473.1589G}

We start by discussing some fundamental properties of the
environmental dynamics in section\,\ref{Sect:Characterize}, followed
by a description of models in section\,\ref{Sect:Model}, the numerical
simulations to characterize the parameters of the acquired \jumbos\,
in section\,\ref{Sect:Results}, and the resulting occurrence rates in
section\,\ref{Sect:Discussion}. We conclude in
section\,\ref{Sect:Conclusions}.

\section{The dynamical characterization of \jumbos}\label{Sect:Characterize}

We initialise our cluster using parameters found by [32], who
numerically modelled disk-size distributions and concluded that the
Trapezium cluster was best reproduced for a cluster containing $\sim
2500$ stars with a total mass of $\sim 900$\,\MSun\, and a half-mass
radius of $\sim 0.5$\,pc. The results were inconsistent with a Plummer
\cite{1911MNRAS..71..460P} distribution, but match the observations if
the initial cluster density distribution represented a fractal
dimension of 1.6 which we adopt here (see \cite{2004A&A...413..929G}).
For consistency with earlier studies, we also perform our analysis for
Plummer models.

Adopting a Plummer distribution of the Trapezium cluster (with virial
radius $r_{\rm vir} = 0.5$\,pc), the cluster core radius becomes $r_c \simeq
0.64r_{\rm vir} \sim 0.32$\,pc with a core mass of 250\,\MSun. This results 
in a velocity dispersion of $v_{\rm disp} \equiv GM/(r_c^2 +
r_{\rm vir}^2)^{1/2} \simeq 0.97$\,km/s. Assuming a mean stellar mass in
the cluster core of 1\,\MSun\, the unit of energy expressed in the
kinematic temperature $kT$ becomes $\sim 8 \cdot 10^{42}$\,erg.

Observed \jumbos\ are found in the mass range of about 0.6\,\MJup\, to
14\,\MJup\, and have a projected separation of 25\,au to $380$\,au.
The corresponding averages are $r_{ij}=200\pm109$\,au, $\langle M_{\rm
  prim}\rangle = 4.73\pm3.48$\,\MJup, and $\langle M_{\rm sec}\rangle
= 2.81\pm2.29$\,\MJup. The median and 25\,\% to 75\,\% percentiles are
$r_{ij} = 193.8^{+78.2}_{-114.1}$\,au $M_{\rm prim} =
3.67^{+1.31}_{-1.57}$\,\MJup, and $M_{\rm sec} = 2.10^{+1.05}_{-1.05}$\,\MJup.
Assuming a thermal distribution in eccentricities, random projection
and arbitrary mean anomalies, we expect the two objects to be bound in
orbits with a typical semi-major axis of $a_{\rm expected} \sim
220$\,au.

To simplify our analysis, let us assume that the observed variation in
projected distances between the two JMOs corresponds to an orbital
separation, and express distances in terms of semi-major axis (see
Appendix\,\ref{Appendix:A} for motivation).  In practice, the
differences between the projected separation and the actual semi-major
axis of the orbit is small. Adopting a statistical approach, a thermal
distribution in eccentricities and a random projection on the sky, the
semi-major axis is statistically $\sim 1.2$ times the projected
separation. Keeping in mind that we do not definitively know whether
the observed \jumbos\, are truly bound, and even if they were, their
underlying eccentricity distribution remains unknown. Nevertheless, in
practice, this difference between projected separation and actual
semi-major axis of a bound population is negligible compared to the
25\% to 75\% uncertainty intervals derived from the simulations.

To first order, the binding energy of \jumbos\ ranges between $\sim
5\cdot 10^{37}$\,erg and $1.4\cdot 10^{41}$\,erg (or at most $\sim
0.02$\,$kT$). This makes them soft upon an encounter with a cluster star. 
On average, soft encounters tend to soften these binaries even further
\cite{1975MNRAS.173..729H}, although an occasional soft encounter with
another planet may actually slightly harden the \jumbo. 

The hardest \jumbo, composed of two 14\,\MJup\, planets in a 25\,au
orbit would be hard for another encountering object of less than
$17\,\MJup$.  For an encountering 1\,\MJup\, object, a 25\,au
orbit would be hard only if the two planets are about three times as
massive as Jupiter.  This implies that \jumbos\, are also generally soft for
any encountering free-floating giant planet unless they are in tight
enough orbits or the perturber is of low enough mass. Overall, independent
of how tight the orbit, \jumbos\, are expected to be relatively short-lived
since they easily dissociate upon a close encounter with any
other cluster member.  The \jumbo\, ionization rate is then determined
by the encounter probability, rather than the encounter parameters.

Once ionized they contribute to the population of free-floating single
objects.  Note that in the Trapezium cluster, even the orbits of 2MASS
J11193254-1137466AB, and WISE 1828+2650 would be soft ($\apgt
0.025$\,kT); they could be the hardest survivors of an underlying
population.

To further understand the dynamics of \jumbos\, in a clustered
environment, and to study the efficiency of the various formation
scenarios we perform direct $N$-body calculations of the Trapezium
star cluster with a population of JMOs in various
initial configurations.

\section{Model calculations}\label{Sect:Model}

For each of our proposed models, ${\cal ISF}$ (in situ formation of
\jumbos\ ), ${\cal SPP}$ (\jumbos\, formed via ejections of a host stars'
outer planets), ${\cal SPM}$ (as planet-moon pairs orbiting a star),
and ${\cal FFC}$ (mutual capture of free-floaters) we perform a
series of $N$-body simulations with properties consistent with the
Trapezium cluster.

Each cluster starts with 2500 single stars taken from a broken
power-law mass-function \cite{2002Sci...295...82K} with masses between
$0.08$\,\MSun\, and $30$\,\MSun\, distributed either in a Plummer
sphere (model Pl) or a fractal distribution with a fractal dimension
of 1.6 (model Fr). All models start in virial equilibrium.  We run
three models for each set of initial conditions, with a virial radius
of 0.25\,pc, 0.5\,pc and 1.0\,pc, called model R025, R050 and R100,
respectively.  We further assume stellar radii to follow the zero-age
main sequence, and the radius of JMOs based on a density consistent
with Jupiter ($\sim 1.3$\,g/cc).

For our proposed models, we initialize a population of single JMOs
and/or \jumbos\, (binary JMOs). The single (and the primaries in
planet pairs) are selected from a power-law mass function between
0.8\,\MJup\, and 14\,\MJup, which is consistent with the observed mass
function \cite{2023arXiv231001231P}. We fitted a power-law to the
primary-planet mass function, which has a slope of $\alpha_{\jumbo}
=-1.2$ (considerably flatter than Salpeter's $\alpha_{\rm Salpeter} =
-2.35$). This choice follows from the observed mass distribution and
is further motivated by the fact that the first dozen discovered
free-floaters having a similarly flat mass function
\cite{2000MNRAS.314..858L}. Indeed, large statistics provided by
gravitational microlensing surveys allowed a reliable measure of the
slope, with $\alpha = -1.3^{+0.3}_{-0.4}$
\cite{2011Natur.473..349S}. This mass function is slightly steeper
than the slope derived for lower-mass ($\aplt 1$\,\MJup) free-floaters
($\alpha = -0.96^{+0.47}_{-0.27}$ \cite{2023AJ....166..108S}).

For each model, we have a special set of initial configurations. The
clusters all have the same statistical representation. But the
distribution of JMOs and \jumbos\, varies per model.  In
figure\,\ref{Fig:models} we sketch the various models.

\subsection{Model ${\cal FFC}$: JMOs as free-floating among the stars}\label{Sect:FFC}

For the models with free-floating JMOs, model ${\cal FFC}$, 
we sprinkle the single objects, with mass taken from a power-law distribution
of slope $\alpha = -1.2$, into the cluster potential as single
objects using the same initial distribution function as we used for
the single stars (either Plummer or fractal).  These models were run
with $\sim 600$\, objects with a mass $>0.8$\,\MJup. 

We performed additional runs with $10^4$ free-floaters. Some of these
runs have a different lower limit to the mass function, to keep the
number of objects with a mass $>0.8$\,\MJup\, at $\sim 600$ (assuming
that lower-mass objects are unobservable).  Each simulation is evolved
for 1\,Myr, after which we study the population of free floating JMOs
and the population of \jumbos. A few simulations were extended to
10\,Myr, to study the long-term survivability of \jumbos.

\subsection{Model ${\cal SPP}$: Star hierarchically orbited by two planets}

For scenario ${\cal SPP}$, 150 stars below and above 0.6\,\MSun\,
are selected to host planetary systems. The mid-mass point (of 0.6\,$\MSun$) 
represents twice the mean stellar mass in the mass function.

As the case for the model described in section\,\ref{Sect:FFC}, the
mass of the primary planet, $M_{\rm prim}$, was chosen from a
power-law distribution with slope $\alpha=-1.2$. The mass of the
secondary planet, $M_{\rm sec}$, was selected randomly from a thermal
distribution between $0.2 M_{\rm prim}$ and $M_{\rm prim}$.  The more
massive planet can therefore either be the inner planet or the outer
one. A consequence of our mass-ratio distribution is that we have a
slight preference for planets of comparable mass (as observed), and
that we have a population of $\aplt 0.8$\,\MJup\, objects. This
low-mass population contributes to $\sim 7.3$\,\% of the total.

The distance from the first
planet $a_1$ and the second planet $a_2$ (such that $a_2>a_1$) are
selected according to various criteria.  The inner orbit $a_1$ was
selected randomly between 25\,au and $400$\,au from a flat
distribution in $a$.  The outer orbit, $a_2$, was typically chosen to
be five times larger than the inner planet's Hill radius.  This
guarantees the stability of the planetary systems if isolated.

Both planetary orbits are approximately circular, with a random eccentricity
from the thermal distribution between circular and $0.02$.  The two
planets orbit the star in a plane with a relative inclination randomly
between $-1^\circ$ and $1^\circ$. The other orbital elements are
randomly taken from their isotropic distributions.  The system's orientation 
in space then gets randomized.  We perform an
additional series of simulations with pre-specified orbital
separations for the two planets $a_1$ and $a_2$, to follow the model
proposed in \cite{2023arXiv231006016W}. The results of these runs are
presented in figure\,\ref{Fig:fjumbos_from_PP}.

\subsection{Model ${\cal SPM}$: Star orbited by a pair of planet-mass objects}

In the ${\cal SPM}$ models we initialize planet pairs (or planet-moon
pairs) in orbit around a star. The masses of the stars, planets and
moons are selected as in the ${\cal SPP}$ model.  The planet-moon
system's orbit was selected from a flat distribution in $a$ between
$25$\,au and $200$\,au, and with an eccentricity from the thermal
distribution with a maximum of $0.02$.

To warrant the stability of the star-planet-moon system, we choose an
orbital separation such that the planet-moon pair stays within 1/3rd
of its Hill radius in orbit around the star. The planet-moon system is
randomly oriented.  These systems tend to be dynamically stable, but
some fraction may be subject to von Zeipel-Lidov-Kozai cycles
\cite{1910AN....183..345V,1962PSS..9..719L,1962AJ.....67..591K}.

With the adopted range of masses and orbital parameters, the
time-scale for a cycle is of the order of a few Myr.  Two JMOs in a
circular 100\,au orbit around a 1\,\MSun\, star would lead to a
circumstellar orbit of $\sim 3490$\,au. The von Zeipel-Lidov-Kozai
cycle period of such a system is $\aplt 1.6$\,Myr, depending on the
eccentricity of the planet-moon system around the star.

\subsection{Model ${\cal ISF}$: JMOs in weakly bound orbits}

Primordial \jumbos\, (model ${\cal ISF}$) are initialized with
semi-major axis following a flat distribution between 25\,au and
$1000$\,au, an eccentricity from the thermal distribution between 0
and 1. The masses are selected as in model ${\cal SPP}$ although we investigate
both the case where the mass-ratio follows a uniform and thermal distribution.  Each system is subsequently
rotated to a random orientation.  The binaries are scattered in the
cluster potential as single objects using the same initial
distribution function as used for the stars. With the arrival of results,
we expand our analysis on this scenario by exploring a more contrived parameter
space (see section \ref{sect:ISF_explored}).

\begin{figure}
\begin{center}
FFC:    \includegraphics[width=0.3\columnwidth]{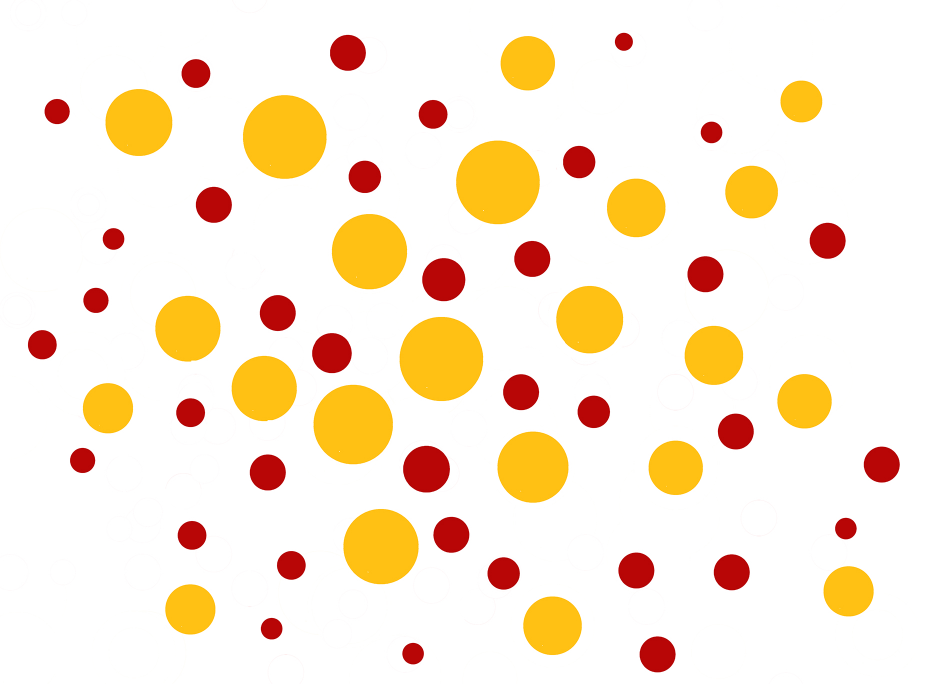}
SPP:    ~\includegraphics[width=0.25\columnwidth,angle=90]{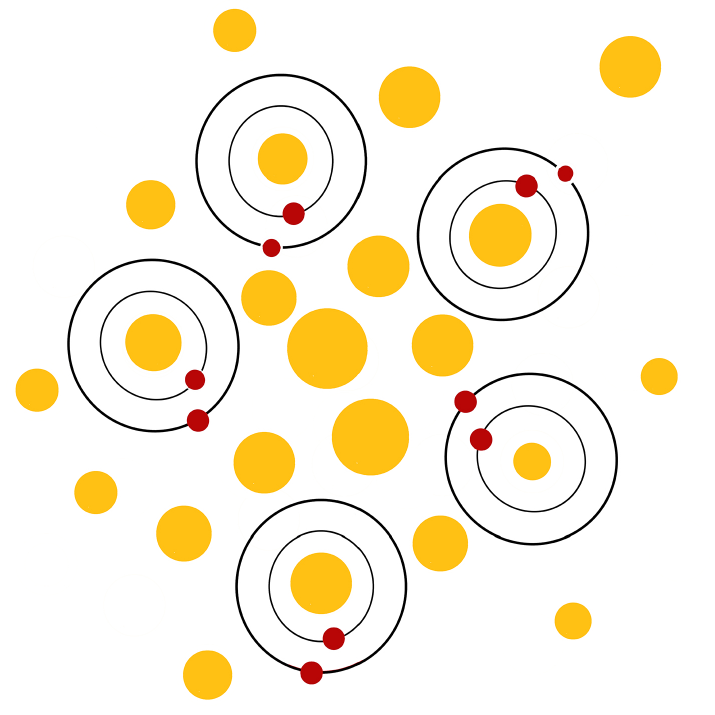}\\
SPM:    \includegraphics[width=0.25\columnwidth]{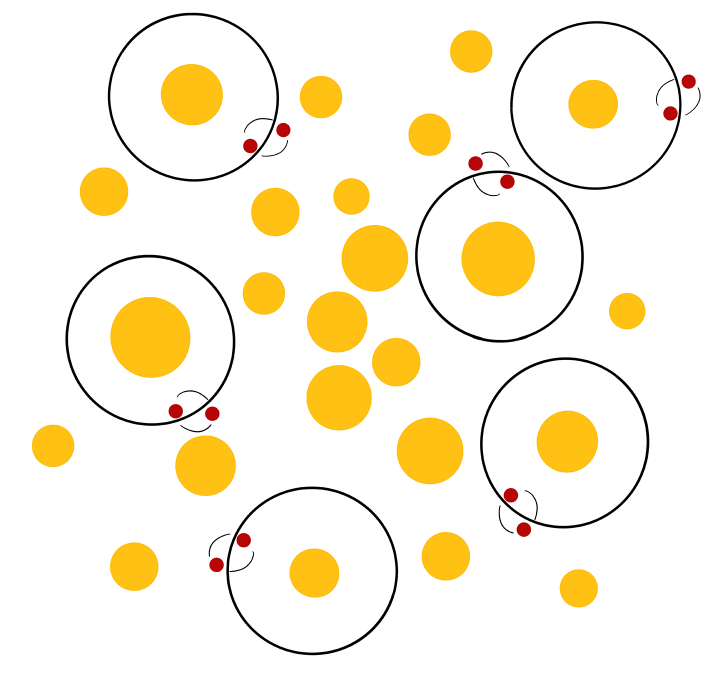}
ISF:    ~\includegraphics[width=0.3\columnwidth]{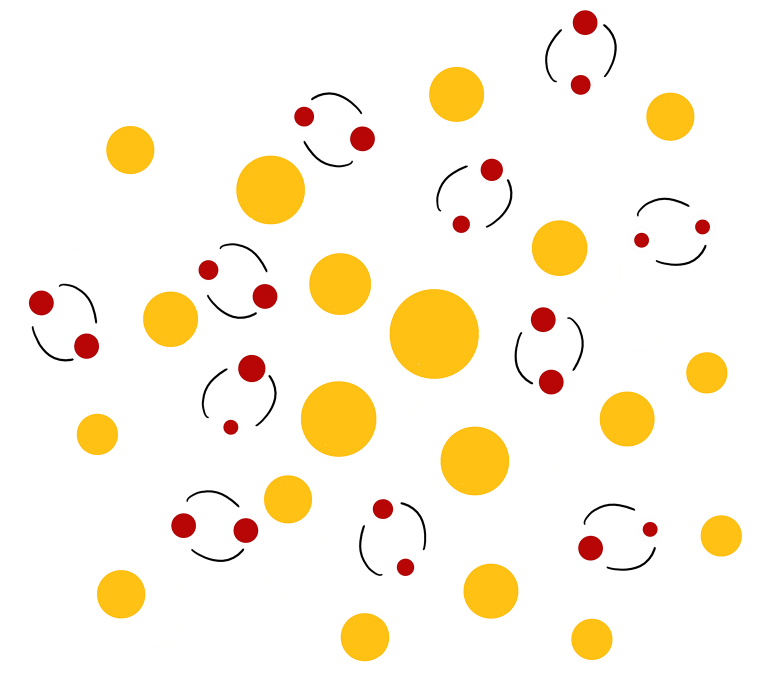}
\caption{Illustration of the four configurations for JMOs in the
  stellar cluster.  Stars are represented with yellow bullets, and
  JMOs in red.  From top left to bottom right we have (as
  indicated): model ${\cal FFC}$ for the free-floating single planets;
  ${\cal SPP}$ for outer orbiting planets; ${\cal SPM}$ as bound
  planet-moon pair orbiting a star, and model ${\cal ISF}$ in situ
  formation of jumbos.}
\label{Fig:models}
\end{center}
\end{figure}

Figure\,\ref{Fig:models} illustrates the four models with a
schematic diagram.

\subsection{The simulations}\label{Sect:Simulations}

All calculations are performed using the 4-th order prediction-correction
direct $N$-body integrator {\sc ph4} \cite{2022A&A...659A..86P}
through the Astrophysical MUltipurpose Software Environment, or AMUSE
for short
\cite{2013CoPhC.183..456P,2013AA...557A..84P,2018araa.book.....P}.
The data files are stored in {\sc AMUSE} formatted particle sets, and
available via zenodo \url{10.5281/zenodo.10149241}; the source
code is available at github \url{https://github.com/spzwart/JuMBOs}.

The script to run the simulations is fairly simple.  It is essentially
the same script from chapter 2 in \cite{2018araa.book.....P},
including the collision-detection stopping condition.  Runs are
performed with the default time-step parameter $\eta=0.03$, which
typically leads to a relative energy error $<10^{-8}$ per step, and
$\aplt 10^{-5}$ at the end of the run. The fractal initial conditions
with small virial radius (0.25\,pc) are, not surprisingly, the most
taxing.  The relative energy error in these runs can be somewhat
higher at times, but never exceed $10^{-3}$, which according to
\cite{2041-8205-785-1-L3}, suffices for a statistically reliable
result.  Snapshots were stored every 0.1\,Myr and we analyse the data
at an age of 1\,Myr.

Although incorporating stellar evolution, general relativity and the
Galactic tidal field would be straightforward in the AMUSE framework,
we decided to ignore those processes.  We do not expect any stars to
effectively lose much mass during the short duration of the
simulations. Meanwhile, incorporating general relativity would have
made the calculations expensive without much astrophysical gain, and
the tidal field hardly has any influence on the close encounter
dynamics in the cluster central portion.  Moreover, in the
observations, \jumbos\, seem to be confined towards the cluster's
central region.  However, this is in part an observational selection
effect. Namely, it is difficult to identify JMOs away from the cluster
center due to intercluster extinction and the lower coverage at the
periphery (S.\, Pearson, private communication).

\subsection{Finding \jumbos}

Considering the average local kinetic energy of surrounding objects
(stars and planets), \jumbos\, are soft, complicating their
identification in the numerical models.  Generally, one considers hard
binary pairs or multiples in direct $N$-body simulations, and finding
soft pairs requires some extra effort.  We search for \jumbos\, by
first finding every individual objects' nearest neighbors, and
determining their binding energy. If an individual's nearest neighbour
has another object identified as its own nearest neighbor, we adopt
that as the close pair, and the initially selected object as a
tertiary. Afterwards, we order the particles in terms of distance and
binding energy, on which the eventual designation is based.

Instead of identifying \jumbos\, as bound pairs, we also analyse the
data only considering nearest neighbors using connected components. With
this method we do not establish \jumbos\, as bound objects, but as
close pairs. This second method mimics observations, in which
boundness cannot yet be established.

We denote single stars $s$, and planets $p$. Pairs of objects are then
placed in parenthesis, for binary stars we write $(s,s)$, a planetary
system with one planet can be $(s,p)$.  A system with two planets then
either becomes $((s,p),p)$, for a hierarchy of planets, or $(s,(p,p))$
for a planet-pair orbiting a star as in the ${\cal SPM}$ model. A
\jumbo\, in this nomenclature becomes $(p,p)$.

\section{Results}\label{Sect:Results}

The main results are presented in table\,\ref{Tab:model_numbers} and
table\,\ref{Tab:orbital_distributions}, but also in the more
fine-tuned simulations presented in
table\,\ref{Tab:Final_ISF_FFC_Results}, and
table\,\ref{Tab:late_formed_jumbos}.

Given the large parameter space available regarding how to distribute
JMO’s among stars, we start by exploring part of this space with a
selection of simulations in which we vary the way JMOs and \jumbos\,
are distributed among the stars. In addition, we cover a small portion
of the cluster parameter space, including the virial radius and the
density profile. All the other parameters we keep constant.

The results of these simulations are reported in
section\,\ref{sect:model_selection}, and in the
tables\,\ref{Tab:model_numbers} and \ref{Tab:orbital_distributions} in
section\,\ref{sect:finetuningbinary_fraction}. Section\,\ref{sect:ISF_explored}
further explores our favorite model, one in which \jumbos\, form as
isolated pairs together with JMOs and stars.

\subsection{Distinguishing between the various models}\label{sect:model_selection}

Table \ref{Tab:model_numbers} summarises the outcomes of our simulations for each scenario.  Rows are named after
their model designation followed by either the letter ``Pl'' for the
Plummer model, or ``Fr'' for the Fractal model.  The model name ends
with the virial radius ``R'' in parsec, here R025 indicates 0.25\,pc,
R050 for 0.5\,pc and R100 for 1\,pc virial radius.

\begin{table}
  \caption{The average number of systems per simulation
    categorized into groups.  The possible outcomes are the number of
    single stars ($n_{s}$), binaries ($n_{(s,s)}$), star orbited by a
    single planet ($n_{(s,p)}$), star orbited by two planets
    ($n_{((s,p),p))}$), single isolated planets ($n_{p}$) and
    \jumbos\, ($n_{(p,p)}$).  Note that we only list those objects
    with a mass $>0.8$\,\MJup.  As a consequence, the total number of
    planetary mass objects do not always add up to 600.  The number of
    stars also do not always add up to 2500 because of collisions
    and hierarchies not listed in the table (see
    section\,\ref{Sect:collisions}).}
 \label{Tab:model_numbers}
 \centering 
 \begin{tabular}{lrrrrrrrrrrrr}
   \hline\hline
   model & $n_{s}$ & $n_{(s,s)}$ & $n_{(s,p)}$ & $n_{((s,p),p))}$ & $n_{p}$ & $n_{(p,p)}$ \\
  \hline
${\cal FFC}$\_Pl\_R025 &  2271 & 87 & 11 & 0 & 580 & 0 \\
${\cal FFC}$\_Pl\_R050 &  2313 & 83 &  3 & 0 & 595 & 0 \\
${\cal FFC}$\_Pl\_R100 &  2331 & 75 &  5 & 0 & 593 & 0 \\
${\cal FFC}$\_Fr\_R025 &  2280 & 93 & 11 & 0 & 584 & 0 \\
${\cal FFC}$\_Fr\_R050 &  2336 & 74 &  4 & 0 & 592 & 0 \\
${\cal FFC}$\_Fr\_R100 &  2312 & 85 &  5 & 0 & 594 & 0 \\
  \hline
  \hline \vspace{-0.75em}\\
${\cal SPP}$\_Pl\_R025 &  2287 &  0 & 84 & 129 & 258 & 0.6 \\
${\cal SPP}$\_Pl\_R050 &  2224 &  1 & 42 & 232 &  93 & 0.7 \\
${\cal SPP}$\_Pl\_R100 &  2204 &  0 & 19 & 277 &  26 & 0.7 \\
${\cal SPP}$\_Fr\_R025 &  2308 & 72 &  8 &   0 & 591 & 0.1 \\  
${\cal SPP}$\_Fr\_R050 &  2279 & 83 & 28 &   6 & 560 & 0.2 \\ 
${\cal SPP}$\_Fr\_R100 &  2327 & 64 & 27 &  10 & 553 & 0.1 \\
  \hline
  \hline \vspace{-0.75em}\\
${\cal SPM}$\_Pl\_R025 &  2480 &  0 & 17 &  3 & 413 & 18 \\
${\cal SPM}$\_Pl\_R050 &  2457 &  0 & 36 &  7 & 341 & 44 \\
${\cal SPM}$\_Pl\_R100 &  2464 &  0 & 22 & 14 & 394 & 15 \\
${\cal SPM}$\_Fr\_R025 &  2320 & 76 &  1 &  0 & 448 &  5 \\ 
${\cal SPM}$\_Fr\_R050 &  2293 & 90 &  2 &  0 & 444 & 17 \\
${\cal SPM}$\_Fr\_R100 &  2361 & 61 &  3 &  0 & 447 & 26 \\
  \hline
  \hline \vspace{-0.75em}\\
${\cal ISF}$\_Pl\_R025 &  2498 & 0 & 0 & 0 & 425 & 23 \\
${\cal ISF}$\_Pl\_R050 &  2498 & 1 & 0 & 0 & 362 & 48 \\
${\cal ISF}$\_Pl\_R100 &  2500 & 0 & 0 & 0 & 246 & 108 \\
${\cal ISF}$\_Fr\_R025 &  2334 & 53 & 7 & 0 & 392 & 0 \\
${\cal ISF}$\_Fr\_R050 &  2309 & 81 & 7 & 0 & 450 & 4 \\
${\cal ISF}$\_Fr\_R100 &  2345 & 73 & 0 & 1 & 454 & 6 \\
  \hline
 \end{tabular}
\end{table}

The ${\cal SPP}$ and ${\cal FFC}$ models systematically fail to
reproduce the observed population of \jumbos\, by a factor of 50 to
400. Changing the initial distribution in the semi-major axis of the inner
orbit from a uniform distribution to a logarithmic distribution
reduces the formation rate of \jumbos\, even further.  There are
several systematic trends in terms of cluster density that depend on 
one's choice of a Plummer or fractal distribution, but it is not clear 
how these models can lead to \jumbos. 

Both models produce a considerable population of
binary stars and single planetary systems, in particular the
fractal distributions, where the typical fraction of dynamically
formed binary stars is around 4\,\%, and the fraction of JMOs
captured by a star is 0.7\,\% per star.  Interestingly, models that
already start with some paired configuration with a star ($\mathcal{SPP}$ 
and $\mathcal{SPM}$) tends to produce more binaries and planetary systems 
than the models where planet-mass objects do not orbit stars ($\mathcal{FFC}$).
The high abundance of hierarchical multiple planets, $((s,p),p)$, in the 
${\cal SPP}$\_Pl and ${\cal SPM}$\_Pl models reflects some of the initial
conditions. These models also tend to produce a relatively rich
population of single planet systems $(s,p)$.

The only models that produce a considerable population of \jumbos\,
are the ${\cal SPM}$ and ${\cal ISF}$ models. The former however
requires the binaries to orbit $a \gtrsim 900$ au, something we deem
infeasible given observations and the size distribution of
circum-tellar disks in the Trapezium cluster. For the latter, the
Plummer distributions tend to produce a sufficient number of \jumbos\,
whereas the fractal model produces too few.

\begin{figure}
    \centering
        \includegraphics[width=0.75\columnwidth]{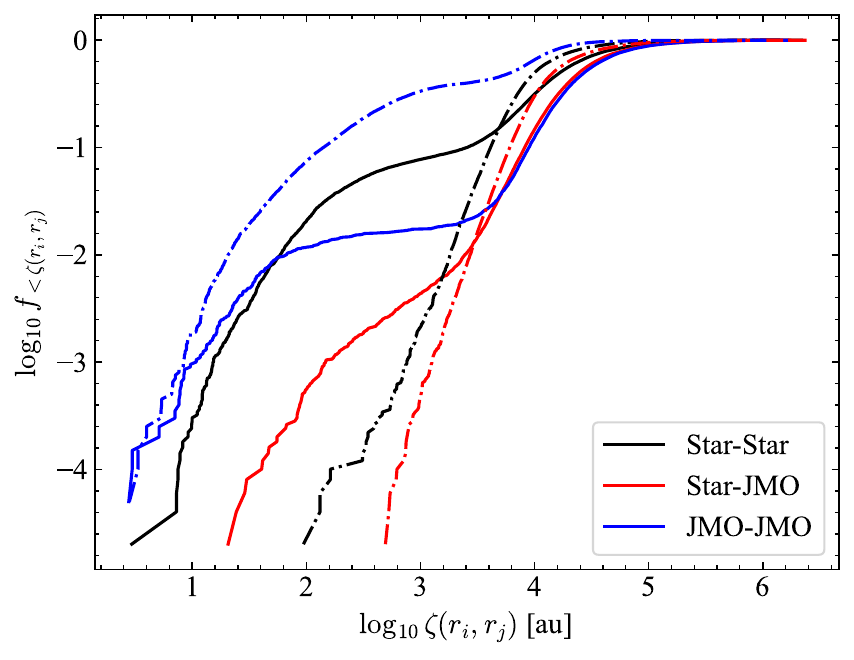}
        \caption{Mutual nearest neighbor distribution function for
          model ${\cal ISF}$\_Fr\_R050 (solid lines) and ${\cal
            ISF}$\_Pl\_R050 (dash-dotted lines). Our analysis only
          considers single objects such that \jumbos\, are
          deconstructed into two JMOs.}
        \label{Fig:twopoint_correlation_ISF_Fr050}
\end{figure}

Figure\,\ref{Fig:twopoint_correlation_ISF_Fr050} shows the cumulative
distribution of the two-point correlation function, $\zeta(r_i,r_j)$,
between stars and JMOs for models ${\cal ISF}$\_Pl\_R050 and ${\cal
  ISF}$\_Fr\_R050.  Note that for calculating the nearest mutual
distance, \jumbos\, are treated as two separate JMO.  This gives rise
to the left shoulder in the blue curve in
figure\,\ref{Fig:twopoint_correlation_ISF_Fr050}. This shoulder is
also visible in the star-star curve (black), whose height exceeds that
of JMO-JMO above $\sim 100$\,au, and falls below within a few tens of
au, where JMO-JMO pairs become more abundant since these correspond to
the tightest initialised \jumbos\, who are able to survive. The
broader and larger JMO-JMO shoulder observed for the Plummer model
results from its ability to better preserve wide \jumbos. The
distribution for the mutual distance between stars and JMOs does not
show such a pronounced shoulder.  Even so, at distances $r_{ij} \apgt
10^4$\,au both distributions converge.  The more violent nature of
fractal distributions is also shown here in two ways. Namely, the
fractal model exhibits a larger proportion of $\lesssim 10^{3}$\,au
detections, implying numerous high-energy encounters. Moreover, the
smaller proportion of JMO-JMO detections within $\lesssim 10^{3}$\,au
highlights the tendency for the fractal model to more efficiently
ionise \jumbos.

Having established the existence of an overpopulation of nearest
neigbors among stars and JMOs, we further explore the orbital
characteristics of the surviving \jumbos.  Table
\ref{Tab:orbital_distributions} lists the median, the 25\% and 75\%
quartiles for the JuMBO, $(p,p)$, distribution in primary mass,
secondary mass, semi-major axis and eccentricity. The table omits
models that produce too few \jumbos\, to calculate the median and
quartiles. We compare observations to the medial values, citing the
quartiles as they better portray the skewness of distributions.  This
is noticeable in the large difference between the range of the lower
and the upper quartiles.

\begin{table*}
  \caption{Simulation results for the models that produce a
    sufficiently large population of \jumbos\, to be considered
    feasible (mainly models ${\cal SPM}$ and ${\cal ISF}$). We present
    the median values and the quartile intervals for 25\,\% and
    75\,\%.  The observed median inter-JMO distance in the Trapezium
    cluster is $r_{ij} = 193.8^{+78.2}_{-114.1}$\,au $M_{\rm prim} =
    3.67^{+1.31}_{-1.57}$\,\MJup, and $M_{\rm sec} =
    2.10^{+1.05}_{-1.05}$\,\MJup\, \cite{2023arXiv231001231P}.  }
\label{Tab:orbital_distributions}
 \centering 
 \begin{tabular}{llllll}
 \hline\hline
model&$\langle M_{\rm prim} \rangle$/\MJup & $\langle M_{\rm sec} \rangle$/\MJup & $\langle a \rangle$/au & $\langle e \rangle$ \\
 \hline \vspace{-0.75em} \\ 
 ${\cal SPM}$\_Pl\_R025 & $3.6^{+1.4}_{-2.7}$ & $1.1^{+0.2}_{-2.1}$ & $99.1^{+37.2}_{-36.4}$ & $0.47^{+0.18}_{-0.13}$ \vspace{0.25em}\\
 ${\cal SPM}$\_Pl\_R050 & $4.0^{+2.1}_{-3.7}$ & $1.7^{+0.7}_{-0.9}$ & $94.3^{+32.8}_{-78.9}$ & $0.33^{+0.19}_{-0.27}$ \vspace{0.25em}\\  
 ${\cal SPM}$\_Pl\_R100 & $6.9^{+4.4}_{-3.1}$ & $1.3^{+0.2}_{-1.2}$ & $141^{+37}_{-11}$     & $0.12^{+0.05}_{-0.27}$ \vspace{0.25em}\\
 ${\cal SPM}$\_Fr\_R025 & $2.7^{+1.8}_{-3.2}$ & $1.9^{+1.1}_{-0.4}$ & $35.6^{+20.6}_{-62.8}$ & $0.80^{+0.10}_{-0.18}$ \vspace{0.25em}\\ 
 ${\cal SPM}$\_Fr\_R050 & $9.6^{+7.0}_{-2.0}$ & $2.0^{+0.7}_{-0.6}$ & $82.8^{+21.7}_{-83.0}$ & $0.56^{+0.29}_{-0.18}$ \vspace{0.25em}\\ 
 ${\cal SPM}$\_Fr\_R100 & $4.6^{+2.6}_{-2.4}$ & $1.8^{+0.6}_{-2.5}$ & $73.2^{+27.2}_{-100.4}$& $0.38^{+0.21}_{-0.13}$ \vspace{0.25em}\\
 \hline \vspace{-0.75em} \\ 
 ${\cal ISF}$\_Pr\_R025 & $8.1^{+3.4}_{-4.5}$ & $2.1^{+1.0}_{-1.7}$ & $112^{+41}_{-182}$     & $0.43^{+0.08}_{-0.19}$ \vspace{0.25em}\\
 ${\cal ISF}$\_Pl\_R050 & $7.0^{+3.3}_{-4.2}$ & $2.0^{+0.7}_{-1.0}$ & $296^{+169}_{-190}$    & $0.62^{+0.19}_{-0.12}$ \vspace{0.25em}\\
 ${\cal ISF}$\_Pl\_R100 & $5.3^{+2.4}_{-3.2}$ & $1.6^{+0.7}_{-1.4}$ & $458^{+265}_{-194}$    & $0.67^{+0.23}_{-0.14}$ \vspace{0.25em}\\
 ${\cal ISF}$\_Fr\_R050 & $4.8^{+3.1}_{-4.4}$ & $1.4^{+0.8}_{-2.4}$ & $37.6^{+15.1}_{-42.5}$  & $0.77^{+0.06}_{-0.05}$ \vspace{0.25em}\\  
 ${\cal ISF}$\_Fr\_R100 &$10.2^{+3.8}_{-2.8}$ & $2.0^{+0.5}_{-2.1}$ & $97.1^{+47.3}_{-189.1}$ & $0.77^{+0.17}_{-0.16}$ \vspace{0.25em}\\
 \hline \vspace{-0.75em} \\ 
 \end{tabular}
\end{table*}

The primary masses produced in the models ${\cal SPM}$ and ${\cal
  ISF}$, tend to be on the high side, but the secondary masses are in
the observed range. ${\cal SPM}$ models tend to lead to orbits that
are too tight. Omitting the orbits with $a<25$\,au \jumbos\, does not
improve the median orbital separation.  In terms of the orbital
separation (or projected distance) the best model seems to be ${\cal
  ISF}$ Plummer with a 0.5\,pc virial radius, but the distributions
are wide, and although the 1\,pc ${\cal ISF}$ fractal model exhibits a
low formation rate, it does reproduce the observed separation
distribution. Meanwhile, models ${\cal ISF}$\_Fr at an age of
$\sim50$\,kyr to $\sim0.2$ Myr compare quite favourably to the
observations (see section\,\ref{Sect:Disc:finetuning}) and could hint
at the idea of \jumbos\ forming later in the cluster evolution when
the system has started to relax.

The eccentricities in model ${\cal SPM}$\_Pl are generally smaller
than in the ${\cal SPM}$\_Fr models. Recall that here, planet-moon
pairs started in nearly circular orbits.  In the fractal models, the
eccentricities are more effectively perturbed and thermalized, whereas
in the Plummer models, this does not happen.  In model ${\cal ISF}$,
\jumbos\, start with higher average eccentricities, in which case the
difference in eccentricity between the Plummer and fractal models is
less pronounced. Later, in section\,\ref{sect:ISF_explored}, we also
explore ${\cal ISF}$ models with initially circular orbits, and models
including a population of free-floating JMOs alongside the \jumbos\,.

\subsection{Stellar and planetary collisions}\label{Sect:collisions}

\begin{figure}
\centering
    \includegraphics[width=0.75\columnwidth]{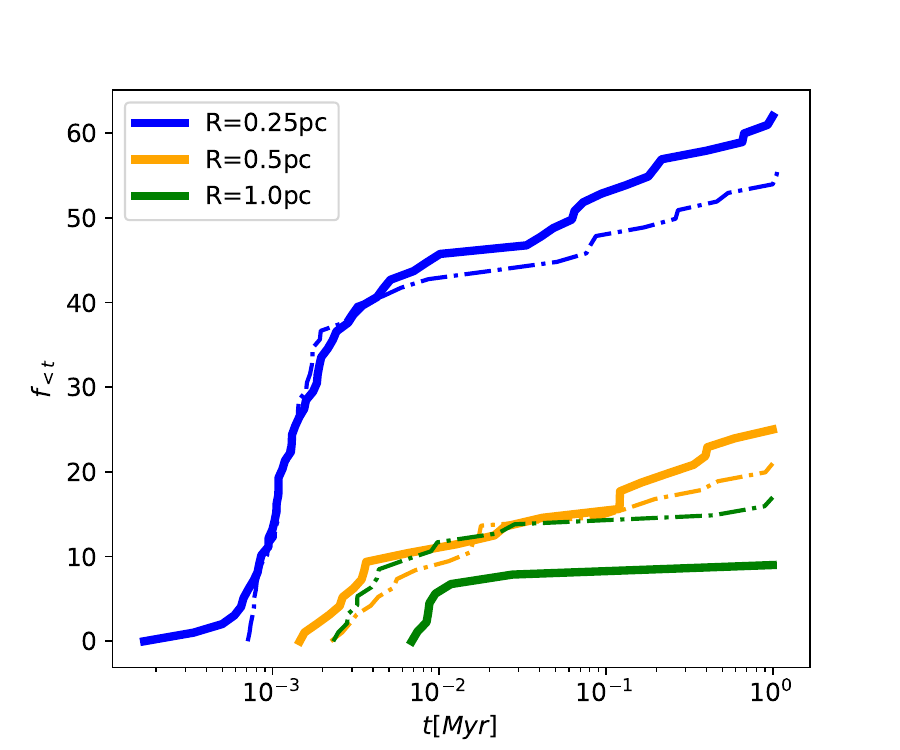
    }
        \caption{Number of collisions as a function of time for model
          ${\cal ISF}\_Fr$ (solid lines) and for models ${\cal
            FFC}\_Pl$ (dash-dotted lines).}
         \label{Fig:collision_evolution_ISF_Fr}
\end{figure}

We encountered several collisions in the simulation. The
majority occur between two stars (83\%), with the rest between a star
and a JMO. Most collisions happen in the fractal models.
In figure\,\ref{Fig:collision_evolution_ISF_Fr}, we present the
cumulative distribution of collisions in the models ${\cal ISF}$\_Fr
with a virial radius of 0.25\,pc (solid blue), 0.5\,pc (orange) and
1.0\,pc (green), and for the equivalent models ${\cal FFC}$\_Pl with the
thin dash-dotted curves.  The higher density naturally leads to more
collisions, which tend to occur at earlier times.  

Plummer models typically yield fewer collisions.  The only Plummer
models in which collisions among stars were common is in model ${\cal
  FFC}$ with 64, 20 and 18 collisions on average per cluster, for
those models with a viral radius of 0.25\,pc, 0.5\,pc and 1.0\,pc,
respectively.  Interestingly, the fractal models from the same series
and virial radii only experience 30, 17 and 14 collisions.  It came as
a bit of a surprise that models ${\cal FFC}$\_Pl lead to so many
collisions, throughout the investigation, no JMO-JMO collisions
occured.

\subsection{Runaway \jumbos}

In addition to mergers, ejection events also occurred.  We adopt the
classical definition of identifying runaway objects if their velocity
with respect to the center of mass of the entire stellar system
(dominated by the bound cluster) exceeds 30\,km/s
\cite{1961BAN....15..265B}.

Not surprisingly, no \jumbo\, escaped the cluster with a high
velocity, but there are some slow runaways, which were born in the
cluster periphery and never experienced an encounter with a nearby
star. Adding a tidal field to our calculations may cause this
population to increase.

Single runaway stars and JMOs are also rare in our simulations. We
attribute this to the lack of hard binaries in the initial
conditions.  We encounter on average one runaway JMO for each of the
fractal models, and typically twice as many runaway stars.  The JMOs,
however, have on average a velocity of $110$\,km/s, whereas the stars
escape with $\sim 63.3$\,km/s.

\subsection{Fine-tuning for the binary fraction and separation distribution}\label{sect:finetuningbinary_fraction}

With results suggesting an ${\cal ISF}$ origin, we further
explore its consequences, and try to derive some of the earlier
\jumbo\, properties to see if those are reconcilable with our
understanding of planet and star formation in
section\,\ref{sect:ISF_explored}.

\begin{figure}
    \centering
    \includegraphics[width=0.49\columnwidth]{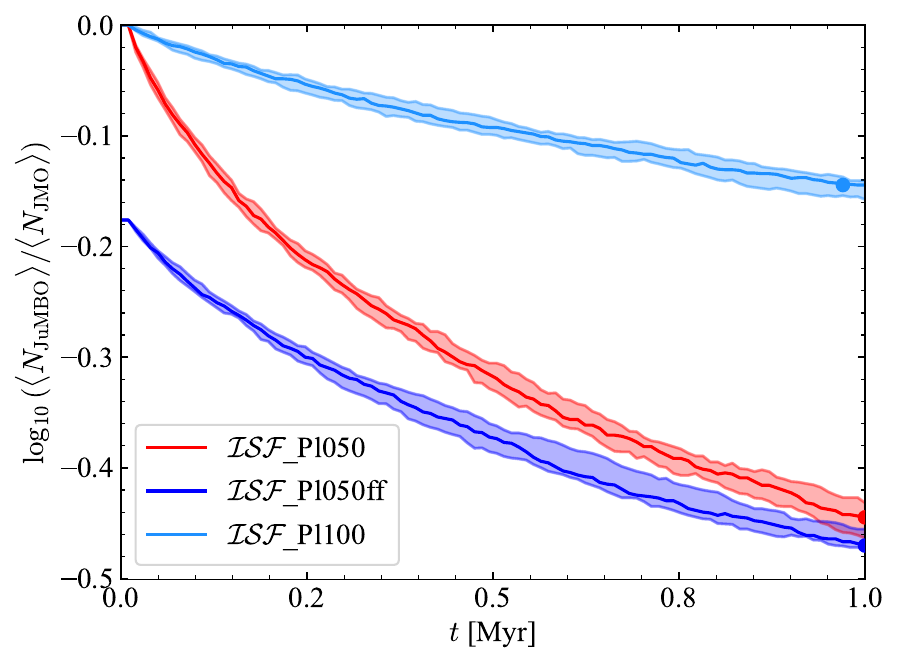}
    \includegraphics[width=0.49\columnwidth]{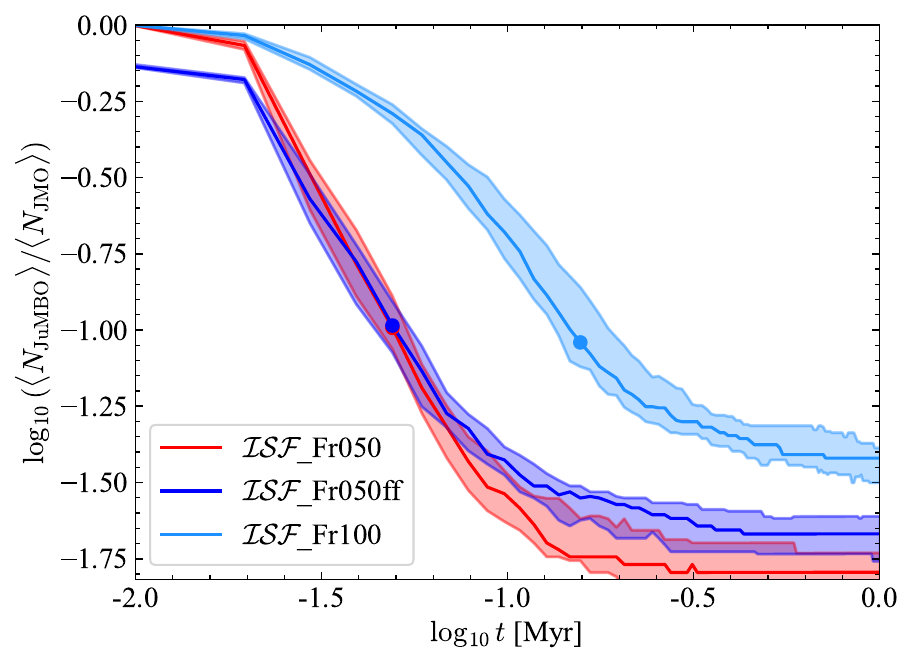}
    \caption{Fraction of \jumbos\ as a function of time for models
      ${\cal ISF}$\_Pl\_R050 and ${\cal ISF}$\_Pl\_R100 (left panel),
      ${\cal ISF}$\_Fr\_R050 and ${\cal ISF}$\_Fr\_R100 (right panel)
      In both cases, we also perform a calculation initialised with a
      population of single JMOs. The number of free-floating objects
      is the same as the number of primordial \jumbos\, and their
      masses are taken from the primary mass function. Scatter points
      denote where in time the model is nearest to the $9\%$ observed
      benchmark for fraction of JuMBOs relative to JMO free-floaters.
    }
        \label{Fig:Fjumbo_vs_time_model_ISF_Pl}
        \label{Fig:Fjumbo_vs_time_model_ISF_Fr}
\end{figure}

Figure\,\ref{Fig:Fjumbo_vs_time_model_ISF_Fr} shows how quickly the
\jumbos\, population decreases in time. The binary
fraction among JMOs initially drops quickly (even
exponentially in the fractal models), before slowing down to a survival 
fraction of ~2\% and ~10\% after 0.2 Myr in the 0.5pc and 1.0pc virial 
radius configurations respectively. For the latter model, the fraction of 
\jumbos\, drops eventually to about 4\,\%; lower than the observed 8\,\%.

\begin{figure}
  \centering
        \includegraphics[width=0.75\columnwidth]{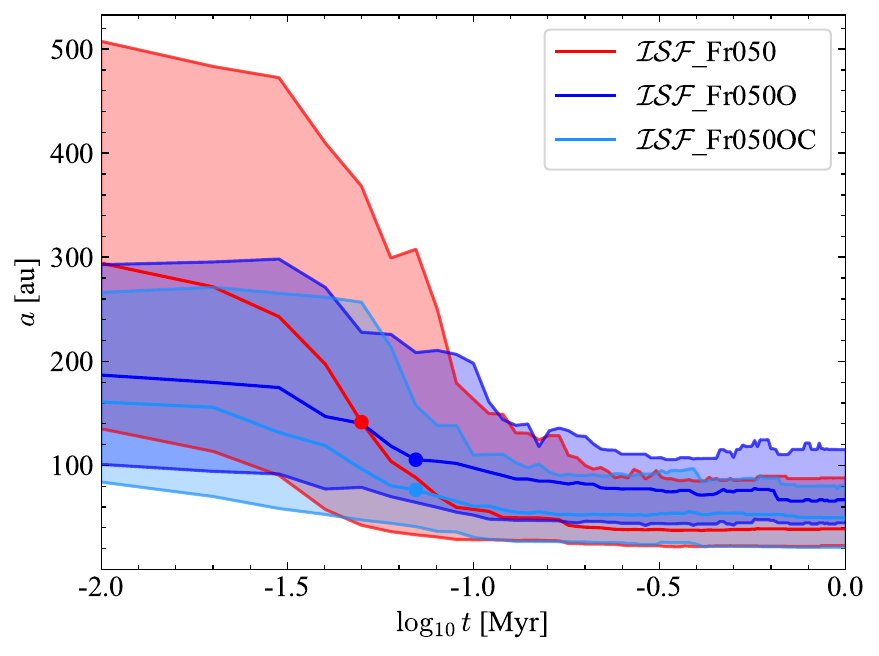}
        \caption{Evolution of the median orbital separation and the
          20\% and 75\% quartiles of the distribution for different
          fractal runs.  Model ${\cal ISF}$\_Fr\_050 looks at the case
          where \jumbos\ are initialised with $0\leq a$ [pc]$\leq
          1000$ while ${\cal ISF}$\_Fr\_050O and ${\cal
            ISF}$\_Fr\_050OC reduce the range to $0\leq a$ [pc]$\leq
          400$. In addition, Fr\_050OC considers initially circular
          JuMBOs, with eccentricities ranging between $0\leq e\leq
          0.2$. }
        \label{Fig:sma_vs_time_model_ISF_Fr}
\end{figure}

Figure \ref{Fig:sma_vs_time_model_ISF_Fr} shows how the orbital
separation of JuMBOs evolves in time and its dependency on the initial
conditions.  Note how quickly the distribution for model ${\cal
  ISF}$\_Fr\_R050 drops with time, before converging in $\aplt
0.1$\,Myr to a median separation $<100$\,au.  In all fractal runs, the
distribution is much narrower and typically has a smaller mean than
the observed properties.

Changing the initial eccentricity and/or semi-major axis distribution have a 
negligible effect on the survival rate of JuMBOs, and is unable to salvage 
the fractal models for explaining the observed population of \jumbos.  

Starting with a tighter population of \jumbos\, will help delay their
ionization, but hardly affects the eventual distribution in orbital
separation.  On the other hand, it will help in making the fraction of
\jumbos\, consistent with the observations
(see\,\ref{Sect:Discussion}). Such consistency is reached $\sim
0.05$\,Myr to $0.2$\,Myr after the start of the simulations.  At this
time, the fraction of \jumbos\, to JMOs, as well as the separation
distribution of the surviving \jumbos\, are consistent with the
observed population. We consider this a strong argument in favor of
\jumbos\, as late formed objects (see\,\ref{Sect:Discussion}).

Although $0.05$ Myr is a blink on cosmic 
scales, and it would be curious why the observed population all formed so near 
in time, one can envisage that \jumbos\ take longer to form. In doing, \jumbos\ 
will emerge after the initial violent phase of a young cluster and exhibit a  
less rapid depletion in their population to that observed in figure  
\ref{Fig:Fjumbo_vs_time_model_ISF_Fr}. It follows that this would provide a  
wider range in time frame with which \jumbos\ may emerge from.

\begin{figure}
    \centering
    \includegraphics[width=\columnwidth]{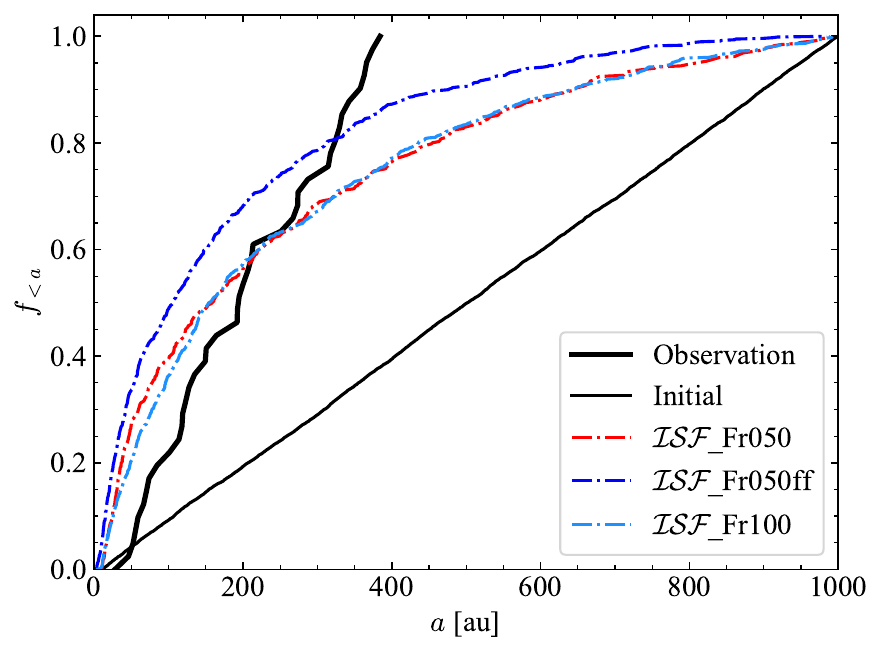}
    \caption{ The cumulative distribution function of JuMBO semi-major
      axis at simulation snapshots where
      $f_{\mathrm{JuMBO}}\equiv{N_{\mathrm{JuMBO}}}/{N_{\mathrm{JMO}}}$
      is nearest to $0.09$. For ${\cal ISF}$\_Fr\_050 and ${\cal
        ISF}$\_Fr\_050FF this corresponds to $\sim 50$ kyr while for
      ${\cal ISF}$\_Fr\_100 this is $\sim0.2$ Myr. In all cases,
      JuMBOs were initialised with a semi-major axis taken from a flat
      distribution ranging between $0\leq a$ [au] $\leq 1000$.  }
        \label{Fig:sma_vs_time_model_ISF_FrA}
\end{figure}

In figure\,\ref{Fig:sma_vs_time_model_ISF_FrA} we present the orbital
distribution for \jumbos\, in model ${\cal ISF}$\_Fr050 (with and
without free-floating JMOs) at an age of 50\,kyr and likewise for ${\cal ISF}$\_Fr100 at an age of $0.2$ Myr. Shortly after the
\jumbos\, are introduced in the simulation, the widest orbits are
being ionized, and by the time the cluster reaches an age of 1.0\,Myr,
the orbital distribution is skewed to values much lower than observations (see figure \ref{Fig:Gen_Semi_Fractal}). At this age, however, the number of \jumbos\, as
well as their orbital separations matches the observed populations in
the Trapezium cluster. 

\subsection{Introducing a population of free-floaters in model ${\cal ISF}$}\label{sect:ISF_explored}

With the current results, we can further constrain the simulation
parameters by repeating several calculations for better statistics and
exploring other parts of the parameter space.  We perform this
analysis for model ${\cal ISF}$ since this model seems most
promising in producing a sufficiently large population of \jumbos\,
in the range of observed parameters.

Once more, we restrict ourselves to a virial radius of 0.5\,pc and
1.0\,pc for both the Plummer and fractal models. Some models
were run with an additional population of single JMO's (denoted with an 
`ff' in the models name). The masses of these free-floaters are 
identical to the \jumbo\, primaries, and they are distributed in the same 
density profile. These models differ from ${\cal FFC}$ since they are 
initialised with a population of \jumbos\,. Each model calculation was 
repeated 10 times to build up a more reliable statistical sample. 
Table \ref{Tab:SF_Res} summarises our results.

Models in the top segment of table \ref{Tab:SF_Res} have more flexible
initial conditions and are initialised with $500$ \jumbos\,, and when
applicable, $500$ free-floating JMOs. The JuMBOs themselves are
initialised with semi-major axis taken from a uniform distribution and
ranging between $0\leq a$ [au]$\leq 1000$. Primary masses are also
taken from a flat distribution with the secondary having masses drawn
from a mass ratio with uniform distribution. In all cases, the mass
ranges between $0.8\leq M\ [\mathrm{M_{Jup}}]\leq 14$.

Unlike the top segment, these `O' models fix the number of JMOs to
$600$, ($n_{\mathrm{\jumbo}} + n_{\mathrm{FF}} = 600$) reflecting the
observations \cite{2023arXiv231001231P}. In cases where the number of
free-floaters, $n_{\mathrm{FF}}$, is not zero, we take into account
the typical survival rate of \jumbos\ after $1$ Myr (first column of
the first segment of table \ref{Tab:SF_Res}) to get the correct
proportion.

Finally, we run an additional series of simulations in which the
primordial \jumbos\, have an eccentricity between 0.0 and 0.2 sampled
from a uniform distribution, rather than the usual thermal
eccentricity distribution (model Fr\_050ffOC where the `C' denotes
circular). Models ending by the letter 'L', such as model FR\_R050ffL,
are those for which we extended the simulation time to 10\,Myr.

\begin{figure}
    \centering
    \includegraphics[width=0.49\columnwidth]{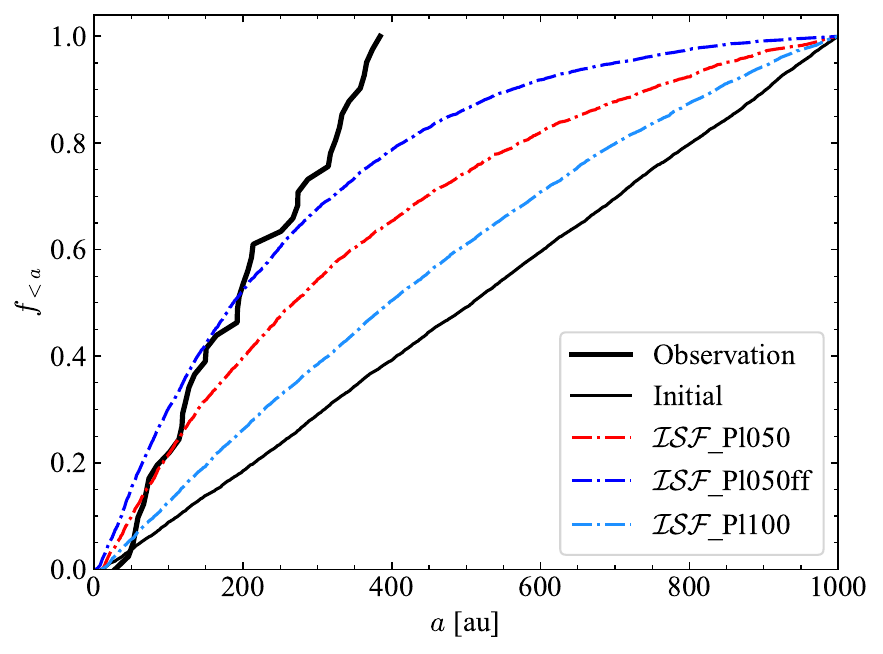}
    \includegraphics[width=0.49\columnwidth]{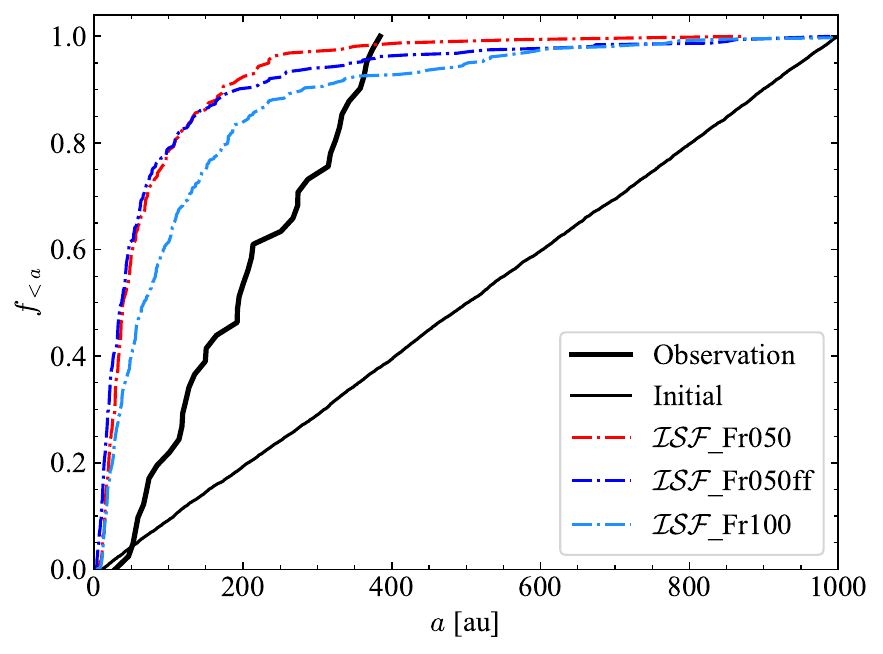}
    \caption{Cumulative distribution of surviving \jumbo\, semi-major
      axis distribution for models Pl\_050, Pl\_050ff, and Pl\_100
      (left panel), and Fr\_050, Fr\_050ff, Fr\_100 (right panel) after $1$ Myr simulation time.}
    \label{Fig:Gen_Semi_Plummer}
    \label{Fig:Gen_Semi_Fractal}
\end{figure}

Overall, we note that although the orbital distribution of the
surviving \jumbos\, in the Plummer models better coincide with
observations than those in the fractal models (see
figure\,\ref{Fig:Gen_Semi_Fractal}), the tail to wider orbits are more
pronounced, potentially highlighting problems with initial conditions.
Reducing the maximum orbital separation from 1000\,au to 400\,au helps
resolve this discrepancy, but it seems a bit unfair to tune the
initial conditions to mimic the observed parameters if the majority of
primordial \jumbos\ survives.  We apply this small but important
change in models in the bottom segment (and identified with an `O' as
their final letter); these models have their JuMBOs initialised with
$25\leq a$ [au]$\leq 400$ and masses drawn from a power-law with a
slope of $\alpha_{\jumbo} =-1.2$ (as was the case for ${\cal PPM}$ and
${\cal SPM}$) and with a thermalised mass ratio.  It would be
relatively straightforward to acquire a satisfactory comparison
between simulations and observations if we start the simulations with
a population of \jumbos\, that reflects the observations.
Since \jumbos\, in Plummer models only marginally evolve, here we
focus mostly on fractal runs which exhibits a natural way of trimming
the wide JuMBO population.
      
\begin{table*}
         \caption{Statistics on the surviving \jumbos. $\langle
           ...\rangle$ gives the median, while the $\pm$ denote 
           the lower and upper quartiles. Col. 1: The fraction of
           \jumbos\, present at the end of the simulation relative to
           the number initialised. Col. 3: The mass ratio of \jumbo\,
           systems. Col. 4: The primary mass of the \jumbo\,
           system. Col. 5: The semi-major axis of the \jumbo\,
           system. Col. 6: The eccentricity of the system.  }
        \label{Tab:SF_Res}
        \centering 
        \begin{tabular}{c c c c c c c c}
        \hline\hline
        Model & $f_{\mathrm{surv}}$ & $\langle M_{\mathrm{prim}} \rangle\ [\mathrm{M_{\mathrm{Jup}}}]$ & $\langle M_{\mathrm{sec}} \rangle\ [\mathrm{M_{\mathrm{Jup}}}]$ & $r_{ij}$ [au] &$\langle a \rangle$ [au] & $\langle e \rangle$\\
        \hline \vspace{-0.75em}\\ 
           Pl\_050     & $0.37^{+0.01}_{-0.02}$ & $8.3^{+2.7}_{-3.1}$ & $3.6^{+2.7}_{-1.8}$ & $233^{+234}_{-137}$ & $268^{+237}_{-152}$ & $0.68^{+0.16}_{-0.22}$ \vspace{0.25em}\\
           Pl\_050ff   & $0.52^{+0.02}_{-0.00}$ & $8.1^{+2.8}_{-3.2}$ & $3.4^{+2.7}_{-1.7}$ & $162^{+167}_{-94}$ & $187^{+176}_{-106}$ & $0.61^{+0.14}_{-0.18}$ \vspace{0.25em}\\
           Pl\_100      & $0.72^{+0.02}_{-0.01}$ & $7.8^{+3.0}_{-3.0}$ & $3.3^{+2.5}_{-1.6}$ & $344^{+271}_{-188}$ & $396^{+250}_{-206}$ & $0.68^{+0.16}_{-0.20}$ \vspace{0.25em}\\
           Fr\_050     & $0.02^{+0.00}_{-0.00}$ & $8.6^{+2.4}_{-3.3}$ & $4.2^{+3.2}_{-1.9}$ & $38^{+52}_{-18}$ & $39^{+50}_{-16}$ & $0.67^{+0.16}_{-0.19}$ \vspace{0.25em}\\
           Fr\_050ff   & $0.04^{+0.00}_{-0.01}$ & $8.8^{+2.7}_{-2.6}$ & $3.9^{+2.5}_{-2.0}$ & $30^{+43}_{-16}$ & $37^{+41}_{-20}$ & $0.62^{+0.14}_{-0.21}$ \vspace{0.25em}\\
           Fr\_100      & $0.04^{+0.00}_{-0.01}$ & $8.3^{+2.4}_{-3.0}$ & $3.8^{+2.4}_{-1.9}$ & $64^{+98}_{-40}$ & $67^{+83}_{-38}$ & $0.68^{+0.16}_{-0.19}$ \vspace{0.25em}\\
           Fr\_050ffL  & $0.03^{+0.00}_{-0.00}$ & $8.1^{+2.9}_{-2.7}$ & $2.7^{+3.1}_{-1.0}$ & $33^{+20}_{-18}$ & $20^{+15}_{-9}$ & $0.61^{+0.20}_{-0.19}$ \vspace{0.25em}\\
           \hline \vspace{-0.75em}\\
           Pl\_050ffO  & $0.76^{+0.01}_{-0.01}$ & $3.7^{+3.5}_{-2.1}$ & $2.1^{+2.5}_{-1.2}$ & $90^{+86}_{-44}$ & $105^{+86}_{-47}$ & $0.61^{+0.15}_{-0.17}$ \vspace{0.25em}\\
           Fr\_050O    & $0.02^{+0.00}_{-0.01}$ & $4.0^{+2.6}_{-1.9}$ & $2.8^{+1.7}_{-1.5}$ & $61^{+46}_{-24}$ & $67^{+48}_{-22}$ & $0.67^{+0.14}_{-0.19}$ \vspace{0.25em}\\
           Fr\_050ffO  & $0.03^{+0.00}_{-0.00}$ & $3.4^{+3.4}_{-1.7}$ & $1.8^{+2.0}_{-0.8}$ & $44^{+37}_{-18}$ & $46^{+38}_{-22}$ & $0.69^{+0.15}_{-0.18}$ \vspace{0.25em}\\
           Fr\_050OC   & $0.02^{+0.01}_{-0.00}$ & $4.7^{+3.2}_{-2.7}$ & $3.6^{+2.6}_{-2.0}$ & $46^{+36}_{-26}$ & $49^{+28}_{-28}$ & $0.45^{+0.33}_{-0.23}$ \vspace{0.25em}\\
           \hline
         \hline                                   
         \label{Tab:Final_ISF_FFC_Results}
        \end{tabular}
     \end{table*}

Adding a population of free-floaters to the Plummer models, makes
considerable difference in the evolution of the \jumbo\, fraction, but
eventually, after about 1\,Myr their fraction converges to roughly the
same value.  They also influence the orbital distribution of \jumbos\,
since the interaction between a relatively tight \jumbo\, and a
relatively low-mass JMO could be hard. This is reflected in the
survival rate of \jumbos. The consequence of this hardening process is
also visible in figure\,\ref{Fig:Gen_Semi_Plummer} where the \jumbos\,
in the models that included free-floaters are, on average,
tighter. Although one can reference to the fact that ionisation will
also lead to a tighter orbital distribution, the enhanced survival
rate between Pl\_050 and Pl\_050ff (and Fr\_050 vs.  Fr\_050ff)
suggests the ionization plays a secondary role to the observed trend.
Although less obvious in fractal runs, we find that JuMBOs are more
likely to survive if simulations include a population of free-floating
JMOs.

This tightening of the orbits establishes itself already at a very early age, as shown in figure\, \ref{Fig:Fjumbo_vs_time_model_ISF_Pl} and figure 
\ref{Fig:sma_vs_time_model_ISF_FrA}.  The \jumbos\,
in these runs, however, remain soft for encounters with any of the
stellar-mass objects in the cluster. The hardening then reduces the
interaction cross-section of the \jumbo\,, making it less vulnerable
to any interaction, including ionization.

The fractal distribution efficiently prunes off any wide orbits since
its violent nature provokes many encounters resulting in
ionization. The tendency for \jumbos\, to ionize at any encounter (JMO
or stellar) is reflected by the little variation between runs of the
same virial radius and the similar survival rates between fractal runs
evolved till $1$ Myr, and Fr\_050ffL which, we recall, evolves until
$10$ Myr.  The ease of ionisation is also reflected by the models with
free-floating JMOs having smaller orbital separations on average.
This trend is even most apparent in the Plummer models (see
figure\,\ref{Fig:Gen_Semi_Plummer}), where the orbital separations are
$\langle a\rangle\sim296$\,au for Pl\_050 compared to $\langle
a\rangle\sim187$ au for model Pl\_050ff.

Overall, we find that free-floating JMOs lead to more \jumbos.  Their
presence tends to yield tighter surviving systems through hardening,
although they also effectively ionize the wider JuMBO systems, this
latter case being no different to any stellar-\jumbo\ interaction.

Increasing the number of JMOs also enhances the chances of two
free-floaters settling into a newly formed binary, on average, only
$0.40$ new \jumbo\, systems emerge in models Fr\_050 compared to the
$0.65$ in Fr\_050ff. For the Plummer models the increased \jumbo\,
formation rate is even more striking by increasing from $1.75$ for
model Pl\_050 to $3.15$ for model Pl\_050ff.  This result sharply
contrasts the ${\cal FFC}$ models, where, irrespective of how many
JMOs were added to the cluster (up to $10^4$), no \jumbos\, formed in any
of our simulations (see section\,\ref{sect:FFC_model_results} for a
discussion). The primordial presence of \jumbos\, mediates their
further formation through interactions with JMOs.

\begin{figure}
    \centering
    \includegraphics[width=0.49\columnwidth]{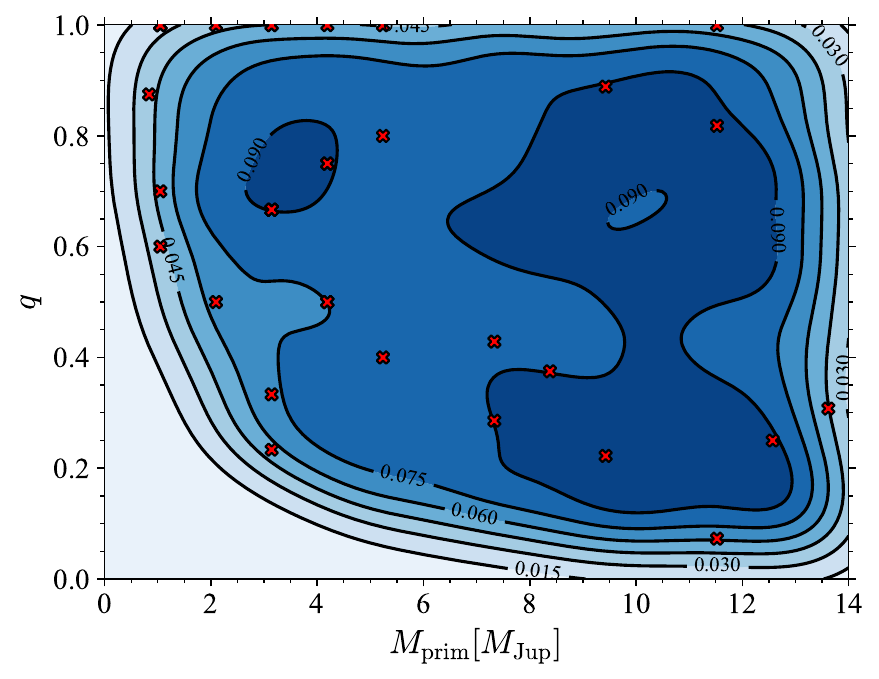}
    \includegraphics[width=0.49\columnwidth]{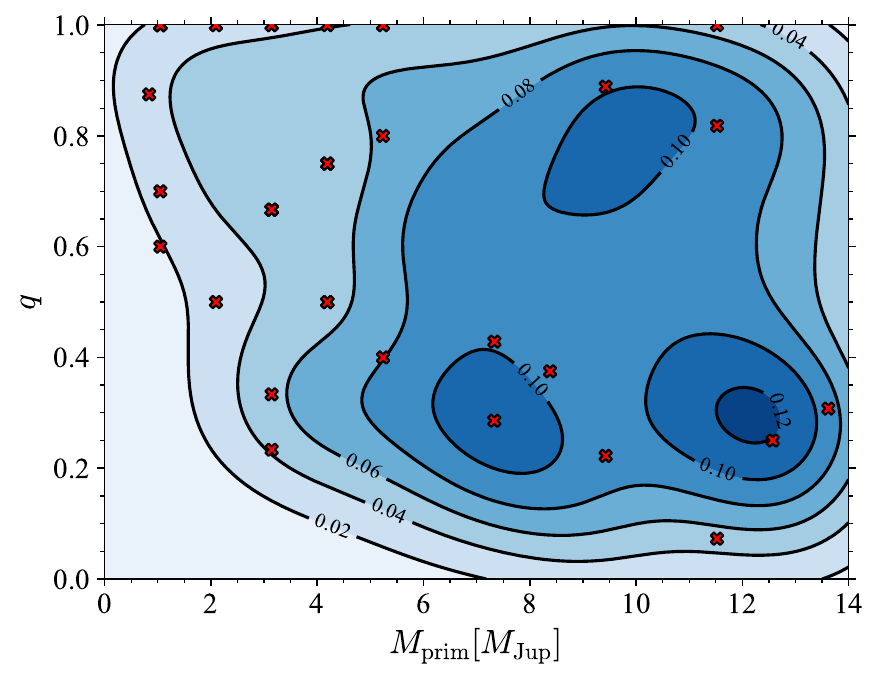}
    \caption{Probability distribution of primary mass versus mass
      ratio for the simulated \jumbos\, in model ${\cal
        ISF}$\_Pl\_R050ff (left) and ${\cal ISF}$\_Fr\_R050ff (right)
      The red crosses show the observed \jumbos.  The primary masses
      for these runs ware generated from a uniform distribution, which
      is still evident in the Plummer model (left), but in the fractal
      model the initial conditions are lost.  }
         \label{Fig:Gen_mdistr_Plummer}
         \label{Fig:Gen_mdistr_Fractal}
\end{figure}

In figure\,\ref{Fig:Gen_mdistr_Plummer} we present the probability
distribution of primary \jumbo\, mass and mass ratio, $q$, for Plummer
and fractal models.  The observed points (red) seem to cover a similar
parameter space, but the overall distribution does not seem to be
consistent with the observations. For the Plummer models this is, in
principle, relatively simple to resolve, by using the observed
distribution as input. For the fractal models such fine-tuning will be
considerably harder because the fraction of survivors is only $\sim
4$\,\%.

\begin{figure}
    \centering
    \includegraphics[width=0.75\columnwidth]{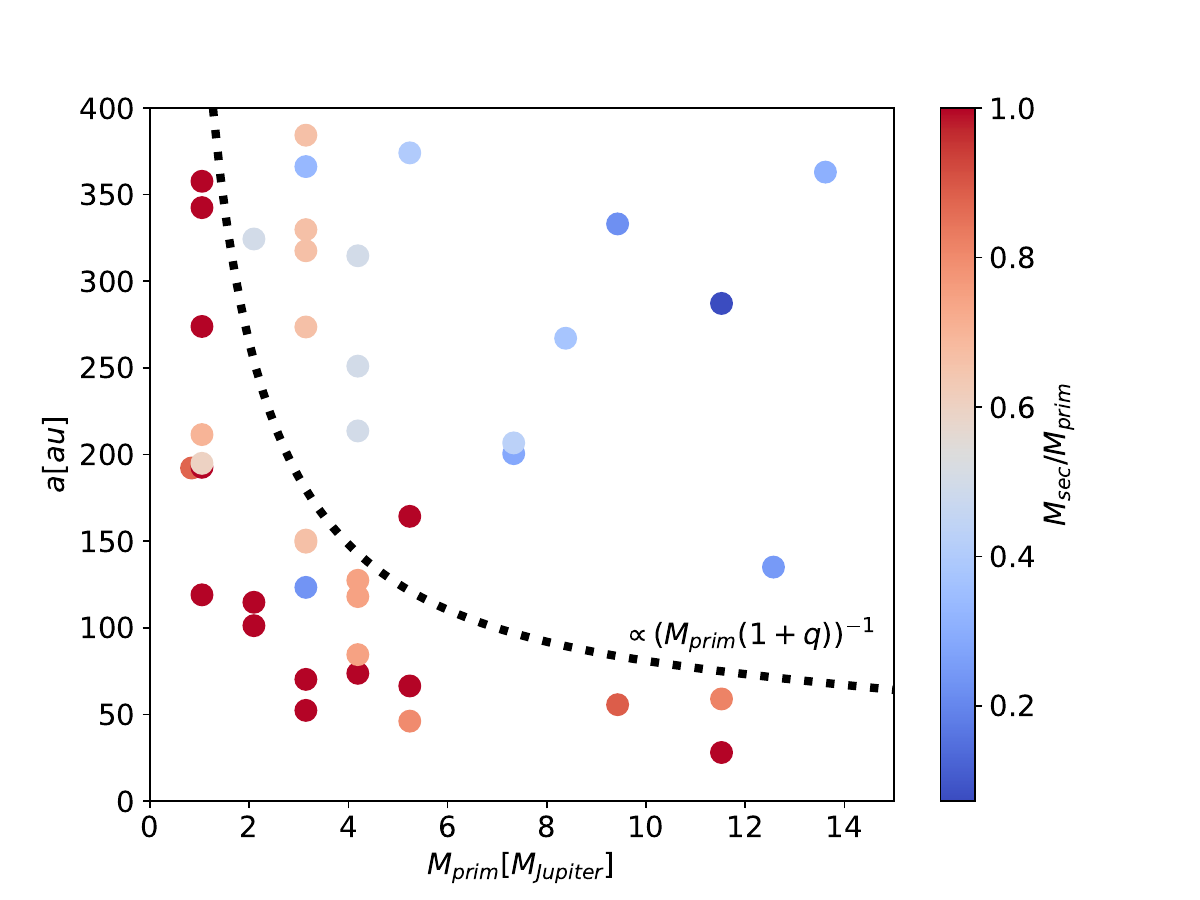}
    \caption{Distribution function of the observed \jumbos, for
      primary mass, observed projected separation and mass-ratio
      (colors). The red bullet points indicate the high mass-ratio
      population, whereas the blue bullets show the low-mass ratio
      population.  The dotted curve separating the two populations
      equals $10^3{\rm au}/(M_{\rm prim}(1+q))$ as indicated.  }
    \label{Fig:obs_q_mprim_sep}
\end{figure}

The overabundance of equal-mass \jumbos\, at $\aplt 4$\,\MJup\,
primary masses in the observations is striking, and hard to
explain. If not just statistics or observational bias, this could
indicate some unexplored formation mechanism. The high-mass-ratio
population \jumbos\, is further illustrated in
figure\,\ref{Fig:obs_q_mprim_sep}, where we plot the observed \jumbo\,
population (data from \,\cite{2023arXiv231001231P}).

Over plotted in figure\,\ref{Fig:obs_q_mprim_sep} is a dotted curve
which represents a separation between the high-mass ratio (red) and
the low-mass ratio (blue) population. We empirically draw this curve
which follows
\begin{equation}
  S(M_{\rm prim}) = 10^3 {\rm au} {1 \MJup \over M_{\rm prim}(1+q)}.
\end{equation}

Here we adopted a distribution in $q$ ranging from $0.2$ to $1.0$
following a thermal distribution such that the lowest mass primary has
an equal-mass secondary, and a ratio of $0.2$ for the most massive
primary. The dotted curve seems to separate high-mass ratio JuMBOs
from the low mass-ratio cases.  One interpretation for this curve is
related to the binding energy, which is proportional to the total
JuMBO mass and inversely proportional to orbital separation.  The
dotted curve then represents a constant binding energy for a $\sim
4.7$\, \MJup\, mass objects in a $a = 1000$\,au orbit around a
$1$\,\MSun-star. Or equivalently, two $10$\,\MJup-objects in a
$25$\,au orbit.

We could continue calibrating the initial conditions for a more
consistent comparison with the observations, but the global parameter
tuning seems to indicate that either the Plummer or the fractal models
with a 0.5\,pc virial radius compare most favorably to the
observations.  At this point, we do not see a natural mechanism to
remove the tail of very wide orbits among the \jumbos, and it is a 
surprising that the observed orbital distribution seems to cut off
rather sharply at about 400\,au.

\begin{figure}
    \centering
    \includegraphics[width=0.75\columnwidth]{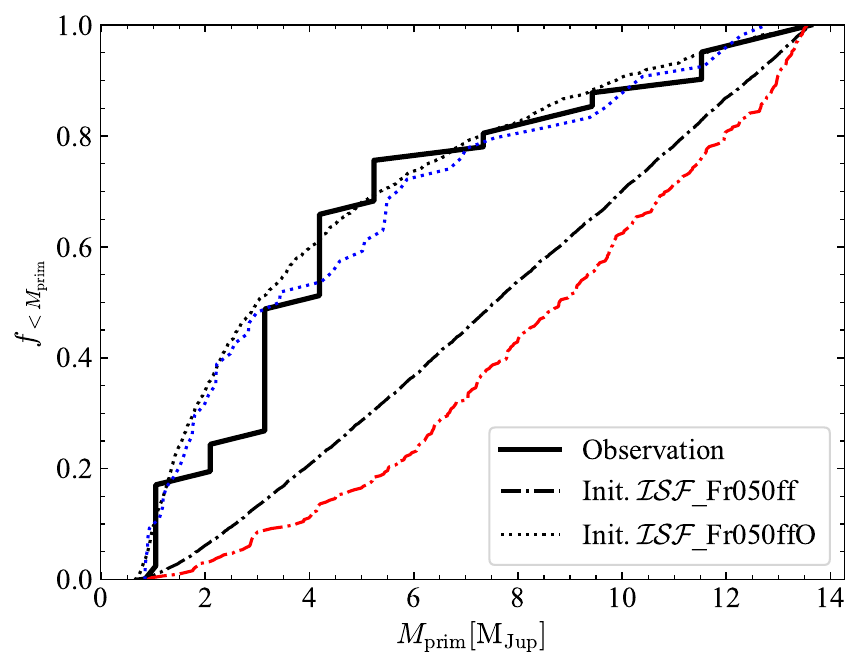}
    \caption{Distribution of the primary mass of the surviving JuMBO
      for models Fr\_050 with an without an adapted mass function. The
      black curves give the initial mass distribution and the colored
      ones the mass distribution after 1\,Myr.}
         \label{Fig:Mdistr_F05}
\end{figure}

In figure \ref{Fig:Mdistr_F05} we show the primary mass function of
the surviving \jumbos\, for two models ${\cal ISF}$\_Fr\_R050.  The
black curves are the initial distributions. Coloured lines of the same
linestyle represent the corresponding final distribution.  The
difference between the initial and final mass primary function is
minimal; one tends to lose some objects in the low-mass end, causing
the curve to become flatter, and the mean mass to increase.  Although
the global trends are similar, the steeper mass function leads to a
better comparison with the observations.  We do not present mass
functions from the other simulations, because the trends are
essentially equivalent, once more highlighting that \jumbos\ are prone
to ionisation no matter with whom they interact.

\section{Discussion}\label{Sect:Discussion}

We explore the possible origin of the rich population of Jupiter-mass
binary objects (\jumbos) in the direction of the Trapezium cluster.
The main problems in explaining the observations hides in the large
number of Jupiter-mass objects (JMOs), their large binary fraction,
and the wide separations. Assuming that they are bound, their orbits
would be soft for any encounter in the cluster, theory dictates that
they should not survive for more than a few hundred kyr ($\sim
0.4$\,Myr for Plummer models, and $\aplt 0.1$\,Myr for the 0.5\,pc
${\cal ISF}$ fractal models).

The ease at which \jumbos\, are ionized is illustrated in
figure\,\ref{Fig:Fjumbo_vs_time_model_ISF_Fr}. It may be clear that
the fractal models, due to their high frequency of strong encounters
in the earliest phase of the cluster lifetime, have difficulty
preserving wide planet-mass pairs. Plummer models are less dynamically
interactive, and the fraction of \jumbos\, remains much higher, where
even relatively wide pairs can survive.

An alternative explanation for the large population of pairs among the
free-floating JMOs might be that they form late.  If \jumbos\, only
formed after $\sim 0.2$\,Myr, they have a better chance of surviving in
the harsh cluster dynamical environment.  Such a late formation would
hardly affect the estimated mass of the objects, because the cooling
curves used to estimate their mass from the observed temperature and
luminosity are roughly flat at such a young age
\cite{2000MNRAS.314..858L}.

The binary fraction continues to drop well after 0.2\,Myr for all
models, and by the time the cluster is 1\,Myr old only $\sim 4$\% of the
binaries in the fractal models survive. The survival fraction in the
Plummer models is considerably higher. The fraction of \jumbo\,
continues to drop, and by the time the cluster is $\sim 10$\,Myr the
fraction of \jumbos\ is $\aplt 2$\,\%.  Interestingly enough,
\cite{2022NatAs...6...89M}, reported the detection of 70 to 170 single
JMOs in Upper Scorpius, which has an age of about 10\,Myr.  None of
the objects in Upper Scorpius is paired, although this observation
could be biased in terms of missing close pairs due to relatively low
resolution of the observations.

Our calculations did not include primordial stellar binaries (or
higher order systems), nor did we take the effect of stellar evolution
and supernovae into account. Those processes may have a profound
effect on the fraction of \jumbos, tending to reduce, rather than
increase their number.

Starting with a large population of ($\apgt 600$) single free-floating
planetary-mass objects among the stars (but without \jumbos\,) would
grossly overproduce the expected number of free-floaters, and
consequently fail to reproduce the observed number of \jumbos\,.  This
model, however, naturally leads to a mass-ratio distribution skewed to
unity, as is observed. We consider this model undesirable by the lack
of a large population of free-floating planets in the Trapezium
cluster and no \jumbos. This could indicate the existence of a large
population of unobservable low-mass objects, but we consider this a
rather exotic possibility.

\subsection{Failure of model ${\cal SPP}$: star with a hierarchical planetary system}\label{Sect:Failure_SPP}

The ${\cal SPP}$ model systematically fails to reproduce the observed
population of \jumbos\, by a factor of $50$ to $400$, leading us to rule out
these scenarios as their possible origins. Changing the initial
distribution in semi-major axis of the inner orbit from a uniform
distribution to a logarithmic distribution reduces the formation rate
of \jumbos\, even further.

To further explore the failure of model ${\cal SPP}$, we perform an
additional series of simulations in the Plummer distribution with
virial radii of 0.25\,pc, and 0.50\,pc.  According to
\cite{2023arXiv231006016W}, the eventual orbital separation of the
\jumbo\, would be consistent with the difference in orbital separation
between the two planets when orbiting the star.  We performed
additional simulations with a mutual separation $a_2-a_1 = 100$\,au and
$a_2-a_1 = 200$\,au, expecting those to lead to consistent results relative to
the observed range in orbital separation for the
\jumbos, as was argued in \cite{2023arXiv231006016W}.  The other
orbital parameters, for the planet masses, their eccentricities and
relative inclinations, in these models were identical to the other
${\cal SPP}$ models.

\begin{figure}
    \centering
        \includegraphics[width=.75\columnwidth]{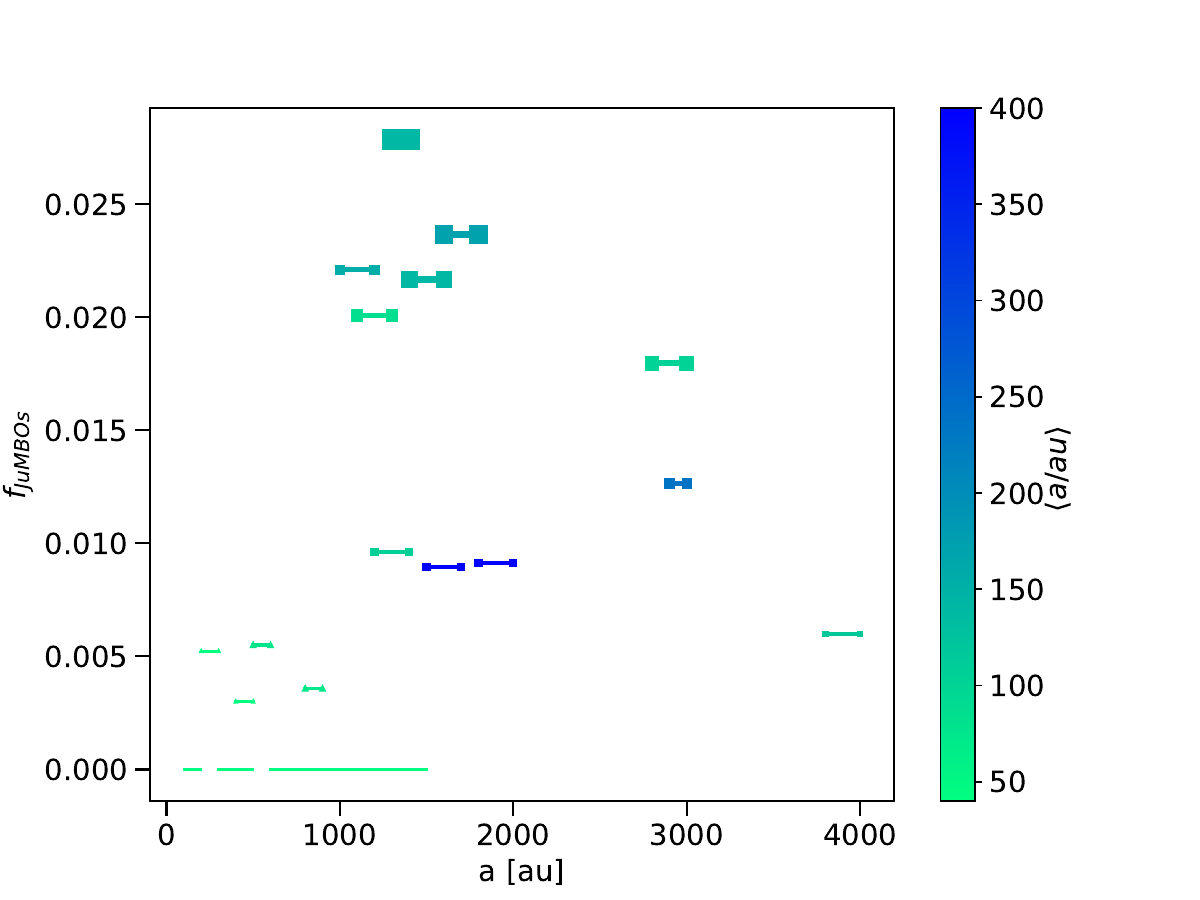}
        \caption{The number of \jumbos\ produced in model ${\cal SPP}$,
          as fraction of the number of free-floating planets for
          various simulations starting with a Plummer sphere and
          virial radius of 0.5\,pc.  The bullet points along each line
          correspond with the adopted orbital separation of the two
          planets ($a_1$ and $a_2$).  The red symbols indicate an
          average orbital separations for the \jumbos\ between 25\,au
          and 380\,au.  The symbol sizes give the number of \jumbos\ ,
          in these simulation linearly scales with a maximum of 9.  }
         \label{Fig:fjumbos_from_PP}
\end{figure}

The results of these simulations are presented in
figure\,\ref{Fig:fjumbos_from_PP}.  The \jumbo-formation efficiency
for these models peaks for an orbital separation $a_1 \apgt 1000$\,au,
but steeply drops for smaller values of $a_1$.  From a total of 45
calculations with various ranges of $a_1$ and $a_2$, 39 produced a
total of 910 \jumbos.  Although results are not inconsistent with
\cite{2023arXiv231006016W}, we find $\sim 50$\,\% wider distribution
in separations than\cite{2023arXiv231006016W}, who argued that the
initial orbital distance $a_2-a_1$ would be preserved.

The JuMBO formation rate is found to be orders of magnitude smaller
than what \cite{2023arXiv231006016W} expected.  They calculate the
rate by means of 4-body scattering experiments, in which a star which
hosts two equal-mass planets that orbit with semi-major axes $a_1$ and
$a_2$ ($a_1 < a_2$), encounters a single star. Their largest
cross-section of roughly $a_1^2$ is obtained if the encounter velocity
$0.8v_\star/v_1$, for an encounter with a star with velocity
$v_\star$. For an encounter at the simulated cluster's velocity
dispersion the inner planet would then have an orbital separation of
$\sim 900$\,au around a 1\,\MSun\, star.

Note that an inner orbital separation of $a_1=900$\,au for a
10\,\MJup\, planet leads to a Hill radius of about 160\,au. An orbit
with $a_2=1100$\,au, then is unstable.  Still, even in the runs where
we use these parameters, the total number of \jumbos\, remains small
compared to the number of free-floaters.  Even if each star in the
Trapezium cluster was born with two such planets at most one-third of
the 42 observed \jumbos\, could conceivably be explained, and the
number of free-floating JMOs would run in the thousands.  A stable
hierarchical system of two Jupiter-mass planets in a circular orbit
around a 1\,\MSun\, star, would be dynamically stable if $a_1 \simeq
120$\,au and $a_2 \simeq 210$\,au (which is hard for an encounter with
a JMO).

The results of the cross-section calculations performed by
\cite{2023arXiv231006016W} are consistent with our direct $N$-body
simulation, but requires initial orbital separation too wide in
comparison with a realistic population of inner-planetary orbits for
JMOs. Indeed, observational constraints on the existence of $\apgt
900$\,au JMOs are quite severe, and we consider it unrealistic to have
300 out of 2500 stars hosting such wide planetary systems. This is
further motivated when one considers the small sizes of the observed
disks are smaller than 400\,au in the Trapezium cluster
\cite{2005A&A...441..195V}.


\subsection{Failure of model ${\cal FFC}$: Free-Floating Jupiter-Mass Objects}\label{sect:FFC_model_results}

In scenario, $\mathcal{FFC}$, we initialize $600$ to $10^4$ single
JMOs in a cluster of stars without \jumbos\, (see
section\,\ref{Sect:FFC}), expecting that some soft pairs form
naturally through interactions with the stars. Soft binary formation,
through three-body interactions is not expected to be very effective
\cite{1976A&A....53..259A}, but with a sufficiently large population,
one might expect a few \jumbos\, to form.

A \jumbo\, can form in models ${\cal FFC}$, when two JMO and a single
star occupy the same phase space volume. In such a scenario, the star
can escape with the excess angular momentum and energy, leaving the
two planetary-mass objects bound.  The distance at which a JMO with
mass $m$ and a star with mass $M$ can be considered bound can be
estimated from the $90^\circ$ turn-around distance, which is $r_{90} =
G(M+m)/v^2$.  For our adopted clusters $r_{90} \simeq
900$\,au. Following \cite{1976A&A....53..259A}, we estimate the
probability of this to happen at $\sim 10^{-2}$ per JMO per relaxation
time. The relaxation time of our Trapezium model cluster is
approximately $t_{\rm rlx} \propto N/(6\ln(N)) t_{\rm cross} \sim
64t_{\rm cross}$. With a crossing time of about 1\,Myr (roughly the
crossing time for our 1\,pc models) we expect $\sim 0.5$ \jumbos\, to
form. The estimates by \cite{1976A&A....53..259A} adopted equal mass
objects, but the more detailed numerical study, carried out by
\cite{2011MNRAS.415.1179M} arrives at a similar number of soft
binaries. The latter study, however, focused on post-core collapsed
clusters, which is not appropriate for our Plummer models, but more in
line with our fractal models. Stellar sub-clumps collapse in the
fractal models within $\sim 0.2$\,Myr, mimicking the post-collapse
evolution as addressed in \cite{2011MNRAS.415.1179M}. Therefore, in
principle, their model is appropriate for our ${\cal FCC}$ models.

\begin{figure}
    \centering
        \includegraphics[width=0.75\columnwidth]{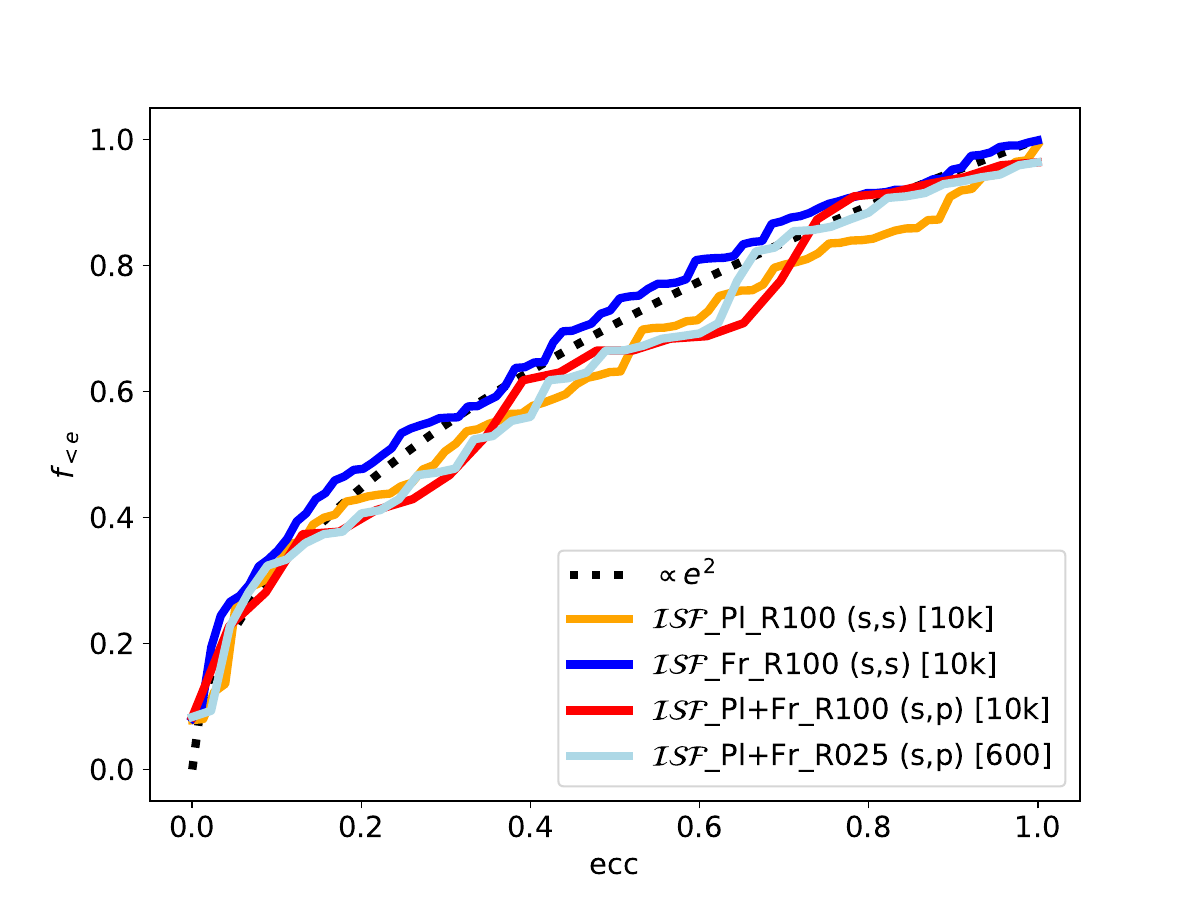}
        \caption{Eccentricity distribution for several models
          (indicated in the lower right inset) for stellar binaries and
          captured planet-mass objects. Overplotted, for comparison,
          is the thermal distribution.}
         \label{Fig:eccentricity_FFC_Fr025}
\end{figure}

Interestingly enough, the ${\cal FFC}$ models produce quite a rich
population of stellar pairs ($82.8\pm 6.6$) and several cases where a
JMO is captured by a star ($6.5\pm3.3$) for the Plummer as well as for
the fractal models. But no \jumbos\, formed. A higher abundance of
stellar pairs compared to planetary captures is somewhat unexpected.
In figure\,\ref{Fig:eccentricity_FFC_Fr025} we present the cumulative
distribution of the eccentricities found in several of our model
calculations. Each of them is consistent with the thermal distribution
(indicated as the black dotted curve).

The secondary masses in the stellar pairs that formed are
statistically indistinguishable from the primary masses of the stars
that captured a planet, and their orbits are wider; $182\pm67$\,au for
the binaries and $288\pm164$\,au for the captured JMOs.  With the
higher masses the binaries are roughly 100 times harder than the JMO
captured systems; straddling the hard-soft boundary.

\subsection{The long-term survivability of \jumbos}

To study the long-term survivability of \jumbos\, we execute $10$ runs
for ${\cal ISF}$\_Fr\_050 until an age of $10$\,Myr.  Our aim is to look 
at the survival of \jumbos\, in older clusters, such as Upper Scoprius. Overall,
the \jumbo\, survival rate decreases rapidly with a half-life
$<1$\,Myr, the survivors have tighter orbits.  The population of
\jumbos\, eventually settles at a population of dynamically hard
pairs, in which case the mean orbital separation $\langle a\rangle
\aplt 20$\,au for two $3\, \mathrm{M_{\mathrm{Jup}}}$ objects. The hardness of
these pairs is mostly the result of the decrease in the cluster
density with time, rather than in the shrinking of the surviving
\jumbos.

\begin{figure}
    \centering
    \includegraphics[width=0.75\columnwidth]{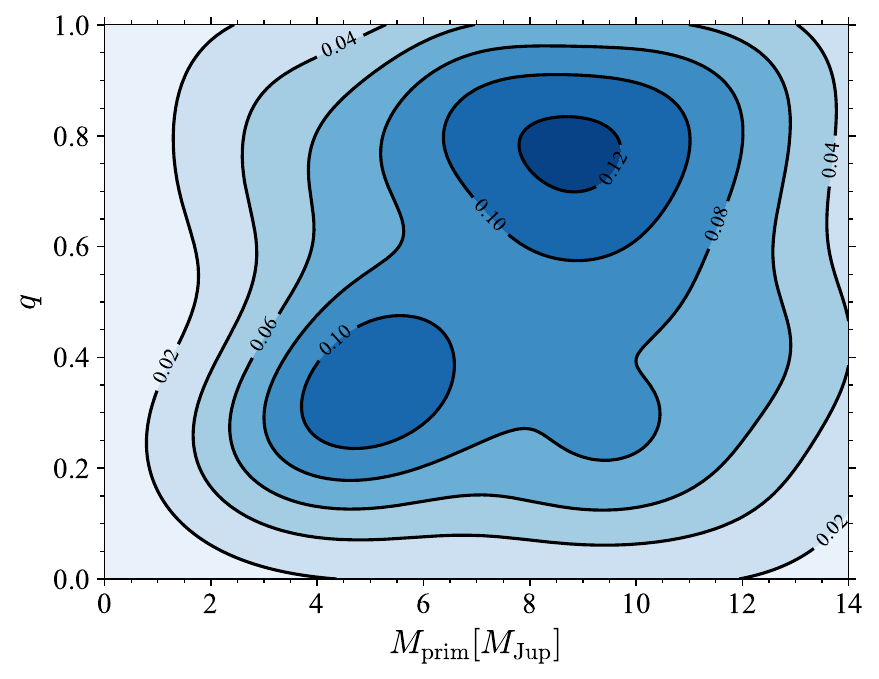}
    \caption{Distribution of primary mass versus the mass ratio for
      model ${\cal ISF}$\_Fr\_050ff at an age of 10\,Myr.  Note that
      here we adopted a uniform mass function
      for the primaries and secondaries scaling their masses through a uniformly sampled
      mass-ratio, $q$.}
         \label{Fig:SimTime_MPrimQ}
\end{figure}

In figure\,\ref{Fig:SimTime_MPrimQ} we present the distribution in
primary mass and mass ratio for simulation ${\cal ISF}$\_Fr\,R050ffL
at an age of 10\,Myr (adopting the uniform primary mass function). An
equivalent diagram for our $1$ Myr runs is shown in the right panel of
figure\,\ref{Fig:Gen_mdistr_Fractal}.  The most likely survivors have
high primary and secondary masses, as could be expected based on the
the hardness of these systems.

\subsection{Observational Constraints}

\begin{figure}
    \centering
    \includegraphics[width=0.75\columnwidth]{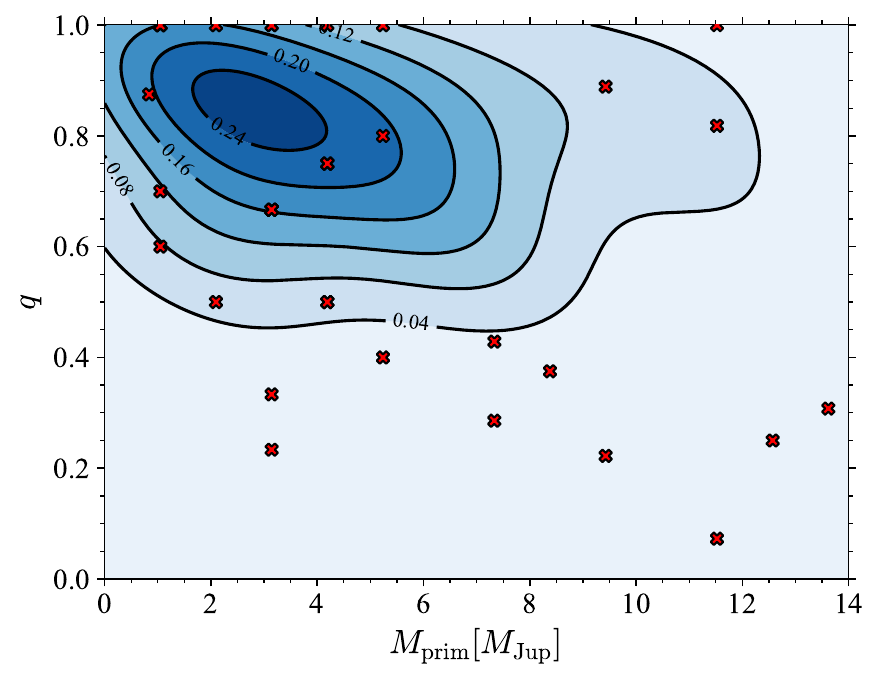}
    \caption{Contours of constant density in the plane of primary mass
      versus mass-ratio for model $\mathcal{ISF}$\_050ff. Red crosses
      denote observed \jumbos.  In this model, the initial mass
      function and mass-ratio distributions for the \jumbos\, were
      skewed to lower mass primaries ($\alpha = -1.2$), and to equal
      mass systems ($0.2\leq q\leq 1.0$, sampled from a thermal
      distribution).  }
         \label{Fig:FractalObs_mdistr}
   \end{figure}

No matter the initial conditions, $f_{\mathrm{surv}}$ and $e$ barely
changes, while $a$ shows only marginal differences.  In turn, the
fractal models exhibit a natural tendency for trimming out wide
binaries. Similarly, \jumbos\ are found not to rely on their mass
parameters. Indeed, no matter the distribution masses are drawn from,
the evolution of the primary mass tends to flatten with a very weak
trend favouring larger $\mathrm{M_{\mathrm{prim}}}$ (see figure
\ref{Fig:Mdistr_F05}). This is reflected in table
\ref{Tab:Final_ISF_FFC_Results} since models initialised with low
primary masses (bottom segment) exhibit upper quartile ranges spanning
larger values and the tendency for \jumbos\ to skew rightwards in the
$M_{\mathrm{prim}}$ vs. $q$ diagram shown in figure
\ref{Fig:FractalObs_mdistr}.

Although we aimed to reproduce the primary mass-function and mass
ratio distribution, we clearly under-represent high mass primaries
with a low mass ratio. For the Plummer models it will be relatively
straightforward to reproduce both populations (the high mass ratio, as
well as those with a low mass ratio) since \jumbos\, in these runs are
rarely ionized, but in the fractal models the survival rate is too low
to still reflect the initial conditions.

\begin{figure}
    \centering
    \includegraphics[width=0.75\columnwidth]{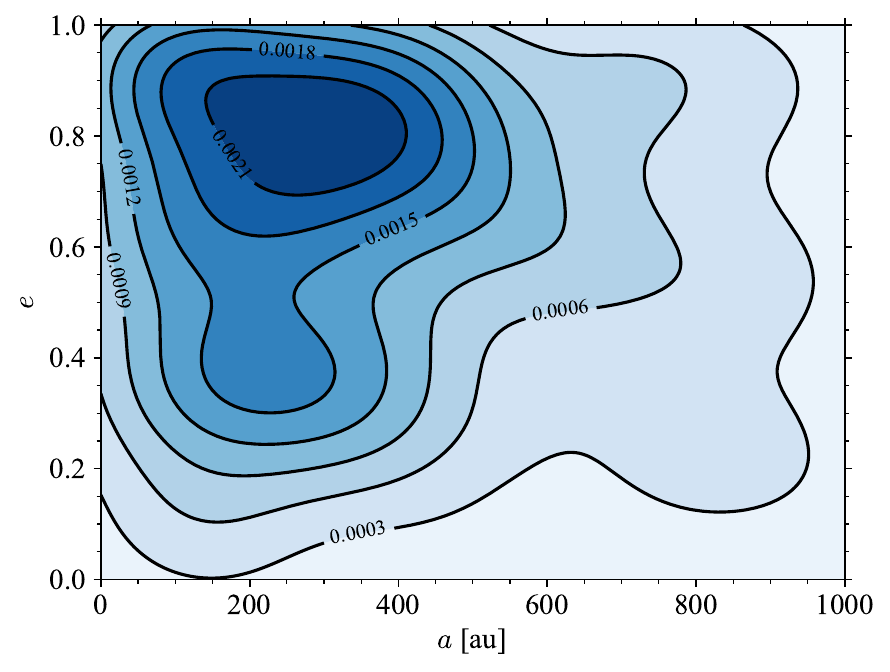}
    \caption{Contours of constant density for semi-major axis and
      eccentricities for JMO-stellar systems for model
      $\mathcal{ISF}$\_Fr\_050ff. These systems are surprisingly
      common in our simulations, and they even tend to be relatively
      tight, though the actual orbits remain dynamically soft.  }
    \label{Fig:MixedSys_OrbParams}
   \end{figure}

In figure \ref{Fig:MixedSys_OrbParams}, we present the distribution of
the dynamically formed JMO-star systems in semi-major axis and eccentricity.
At any given instance in time, a typical fractal run has $\langle N\rangle \sim 
4$ of these systems present in the cluster. The parameter space
is widely covered, with signs of low eccentricity but very wide
($a\geq700$ au) binaries. The vast majority exhibit
large eccentricities and semi-major axis, reflecting their dynamical
origin. The non-negligible amount of these systems emerging provides an
interesting prospect of detecting ultra-cold Jupiters orbiting stars
that have recently fostered them.

Our slight preference for the fractal models stems from their natural
consequence of producing higher-order hierarchical systems, which on
rare occasions produce triple JMO systems (Jupiter Mass Triple
Objects, JuMTOs) as was observed in \cite{2023arXiv231001231P}, and
its capability of removing wide \jumbos.  Contrariwise no triples were
detected in Plummer models and after $1$ Myr.
Nevertheless, if formed in situ,
the Plummer may have a sizable fraction of primordial triples survive:
we are then back at weighing the relative importance of initial
conditions versus dynamical evolution and emergence.


\subsection{Assessment on the origin of \jumbos}

In the ${\cal ISF}$ Plummer models, depending on the clusters
configuration, $\sim 50\%-70$\% of the \jumbos\, are ionized within
1\,Myr, compared to $\sim 96$\,\% for the fractal models.  The
observed population of free-floaters and \jumbos\, can then be
reproduced if the cluster was born with half of its population of JMOs
as free-floaters and the other half as pairs.  The current observed
primary and secondary masses of \jumbos\, would reflect the conditions
at birth, while the semi-major axis and eccentricity distributions
would have been affected considerably by encounters with other cluster
members. These processes tend to drive the eccentricity distribution
to resemble the thermal distribution (probably with an excess of
$\apgt 0.7$ eccentricities \cite{2000IJoMP...15..4871P}). The
semi-major axes of the \jumbos\ would have widened, on average by
approximately 5\% due to encounters with free-floating planet-mass
objects.

Alternative to a Plummer initial stellar distribution we consider
fractal distributions, which are also able to satisfactorily reproduce
the observed populations. In the fractal models, $\apgt 90$\,\% of the
primordial \jumbos\, become ionized, and in principle the entire
observed populations of free-floating JMOs and
\jumbos\, can be explained by a 100\% initial binarity among the
\jumbos. We then conclude that JMOs are
preferentially born in pars with a rather flat distribution in orbital
separations with a maximum of $\sim 400$\,au.  Higher order
multiplicity (JuMTOs and JuMQOs) form naturally from interactions between
two or more \jumbos\, in the fractal models.

This model (${\cal ISF}$\_Fr\_R050) satisfactorily explains the
observed orbital separation distribution, with a $\sim 15$\% excess of
systems with a separation $\apgt 400$\,au. We do not expect a rich
population of orbits with separation smaller than the observed
25\,au. Indeed, when processing results we found that fractal models
whose \jumbos\ are initialised with $10\leq a$ [au] $\leq 1000$,
$\sim65\%$ of \jumbos\ have orbital separations larger than $25$\,au
($\sim90\%$ when we restrict the initial semi-major axis to $25\leq a$
[au]$\leq 400$). This increases to $\sim 90\%$ for Plummer models.
The fraction of pairs among the wide systems is already high, and
there are not enough single JMOs observed to accommodate this tight
binary population, unless a considerable fraction of the single
observed JMOs are in fact such tight binaries.

We have a slight preference for the ${\cal ISF}$ fractal models with
0.5\,pc virial radius because hierarchical triple JMOs form naturally
in roughly the observed proportion (on average $\sim 4$ triples among
$\sim 40$ pairs and $\sim 500$ single JMOs). The singles then
originate from broken-up pairs, and the triples form in interactions
between two JMO pairs.  The dynamical formation of soft triple JMOs is
quite remarkable, and observational follow-up would be of considerable
interest.

The mass function of single JMOs should resemble the combined mass functions 
of the primary and secondary masses of the \jumbos\, since the ionization 
probability for a \jumbo\, does not depend on its mass, but instead on the 
chance encounter with a star. In this sense, the majority of JMOs would originate
from ionized \jumbos.

\subsection{Fine tuning of model ${\cal ISF}$\_R050}\label{Sect:Disc:finetuning}

The short-periods and the small binary fraction of the fractal models
could be salvaged if the \jumbos\, form late. If the majority of the
observed population formed $\apgt 0.2$\,Myr later than the stars, the
cluster's density profile would already have been smoothed out,
leading to fewer strong dynamical encounters. The cluster would
somewhat resemble a \jumbo -friendly Plummer-like structure. Not only
would that mediate the survival of \jumbos\, but it also would allow
them to preserve their orbital characteristics.

In table\,\ref{Tab:late_formed_jumbos} we present a small experiment
to support the argument of \jumbos\, forming late.  The table shows the
fraction of surviving \jumbos, and their orbital parameters. We omitted
the eccentricities, as they are consistent with the thermal
distributon.

The experiment starts with simulation model ${\cal ISF}$\_Fr\_R050.
Instead of running with \jumbos\, we run with 900\, test masses
distributed in the same fractal structure as the stars.  At various
moments in time, we stop the run, and restart it with \jumbos\, and
JMOs.  Each restart is initialized with 300\,\jumbos\, and 600\,JMOs.
The \jumbos\, are taken from the same distribution functions for
primary mass, secondary mass and orbital parameters from the models
${\cal ISF}$; Their maximum orbital separation is 400 au.

\begin{table}
  \caption{Final distribution of fraction of \jumbos\, and orbital
    parameters for models ${\cal ISF}$\_Fr\_R050 but with the time of
    birth for the \jumbos\, at $t_{\rm birth}$.  The best combination
    of values of survival fraction, and orbital separation are
    achieved if \jumbos\, are initialized somewhat later than the
    stars by between 0.2\,Myr or 0.4\,Myr.  The initial fraction of
    \jumbos\, over JMOs is typically 50\,\%.  }
 \label{Tab:late_formed_jumbos}
 \centering 
 \begin{tabular}{lrrrrrrrrrrrr}
   \hline\hline
 $t_{\rm birth}$/Myr & $f_{(p,p)}$  & $\langle M_{\rm
  prim} \rangle$/\MJup & $\langle M_{\rm sec} \rangle$/\MJup & $\langle a \rangle$/au \\
 0.0 & 0.05 &  $8.1^{+3.4}_{-4.5}$ & $2.1^{+1.0}_{-1.7}$ & $112^{+41}_{-182}$ \\
 0.1 & 0.15 &  $5.5^{+2.6}_{-3.5}$ & $2.1^{+1.1}_{-1.3}$ & $156^{+82}_{-95}$ \\
 0.2 & 0.22 &  $5.2^{+2.4}_{-3.4}$ & $1.5^{+0.6}_{-1.2}$ & $192^{+90}_{-92}$\\
 0.4 & 0.24 &  $4.6^{+2.0}_{-4.5}$ & $1.4^{+0.5}_{-1.3}$ & $220^{+116}_{-105}$\\
 0.6 & 0.31 &  $5.0^{+2.3}_{-3.7}$ & $1.5^{+0.5}_{-1.2}$ & $199^{+80}_{-91}$ \\
 0.8 & 0.33 &  $4.4^{+1.9}_{-3.3}$ & $1.3^{+0.5}_{-1.4}$ & $187^{+78}_{-99}$ \\
  \hline
 \end{tabular}
\end{table}

\begin{figure}
    \centering
    \includegraphics[width=0.75\columnwidth]{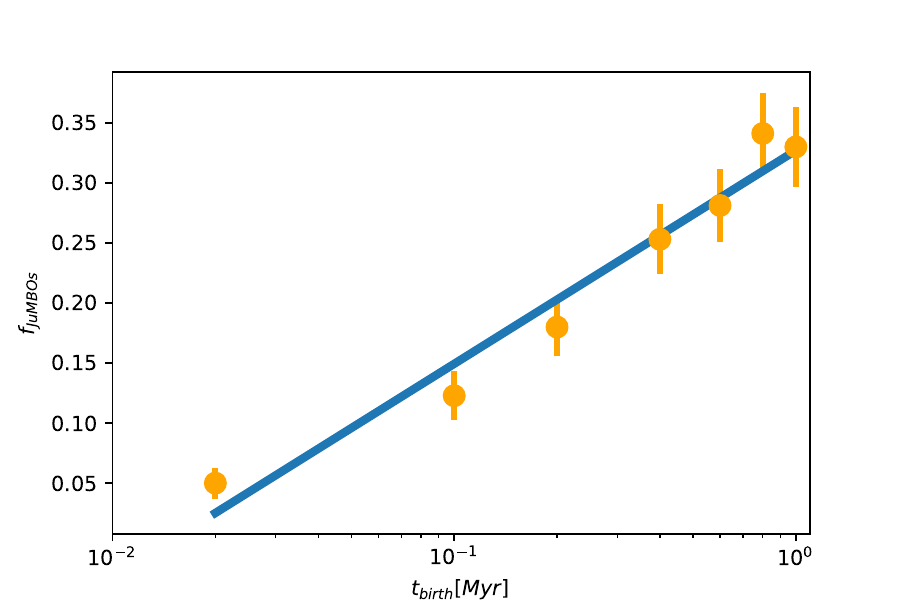}
    \caption{Surviving fraction of \jumbos\, for model ${\cal
        ISF}$\_Fr\_R050 as a function of the moment they were
      introduced in the simulation ($t_{\rm birth}$).  The orange
      bullets present the actual measurements, and the curve is a
      least squares fit, represented by $f_{\jumbos} \simeq 0.33 +
      0.18\log_{10}(t_{\rm birth}/{\rm Myr})$.}
    \label{Fig:survival_fraction_ISF_late_foramtion}
\end{figure}

It should not come as a suprise that the fraction of surviving
\jumbos\, increases when they form later. We illustate this further in
figure\,\ref{Fig:survival_fraction_ISF_late_foramtion}, where we plot
the surviving fraction of \jumbos\, as a function of the time they
were introduced in the simulation.  We started introducing them at the
same time as the stars (left, $t_{\rm birth} = 0$\,Myr), and as last
point they were introduced together with the stars (right most point
at $t_{\rm birth} = 1$\,Myr).  The left-most point essentially
replicates our earlier ${\cal ISF}$\_Fr configurations, and the
right-most point reflects the initial conditions.

\begin{figure}
    \centering
    \includegraphics[width=0.75\columnwidth]{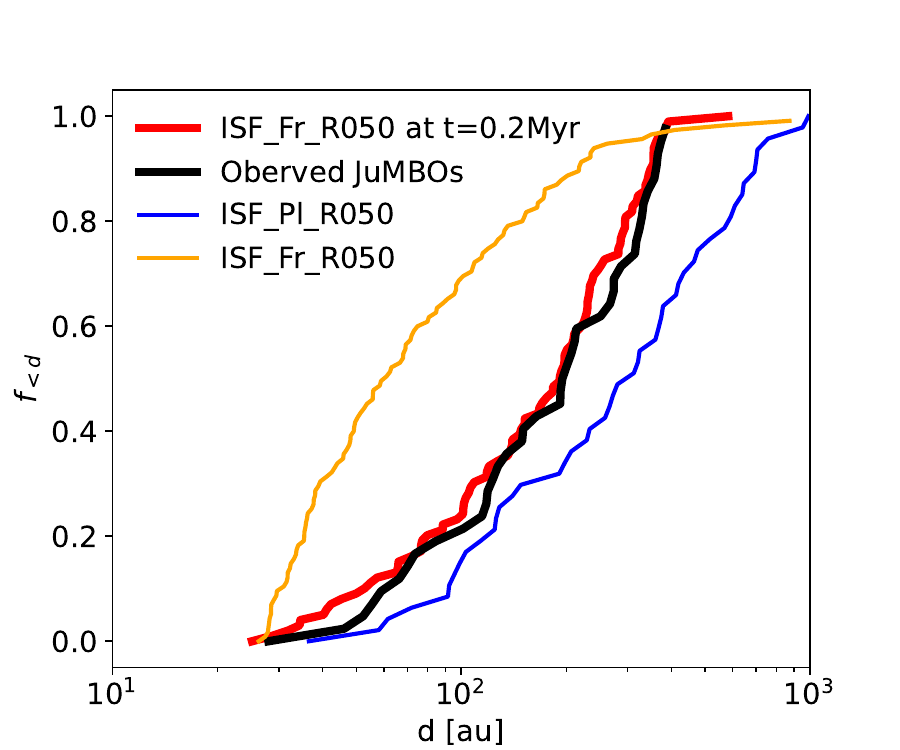}
    \caption{Distribution of orbital separation for the observed
      \jumbos\, (black) and those from the two models ${\cal
        ISF}$\_Fr\_R050 with $t_{\rm birth} = 0.2$\,Myr (red), and
      adopting $t_{\rm birth} = 0$\,Myr we plot both ${\cal
        ISF}$\_Pl\_R050 (blue) and ${\cal ISF}$\_Fr\_R050. The fractal
      model compares well with the observed distribution (KS p-value
      $= 0.939$) compared to the Plummer and fractals distributions
      (KS p-value$<0.003$). The blue and orange curves are also
      presented in figure\,\ref{Fig:Gen_Semi_Plummer}, as the red
      dashed curves (but now in $\log$-scale).}
    \label{Fig:orbital_separation_ISF_late_foramtion}
\end{figure}

In figure\,\ref{Fig:orbital_separation_ISF_late_foramtion}, we present
the orbital distribution for $t_{\rm birth} = 0.2$ of model ${\cal
  ISF}$\_Fr\_R050 (red), and for model ${\cal ISF}$\_Pl\_R050 for
which the \jumbos\, were initialized at birth ($t_{\rm birth} =
0$). The black curve shows the observed projected separation
distribution of the \jumbos\, from \cite{2023arXiv231001231P}.

The consistency of the orbital separation in the fractal model in
figure\,\ref{Fig:orbital_separation_ISF_late_foramtion} is in part by
construction, as this was also the input initial condition.  But
similar input parameters were adopted for the Plummer model and the
fractal model, but in both these cases the \jumbos\, were initialized
together with the stars.


\section{Conclusions}\label{Sect:Conclusions}

The discovery of 40 relatively wide pairs, two triples and 540 single
JMOs in the Trapezium cluster emphasizes our limited understanding of
low-mass star and high-mass planet formation. To derive
characteristics for their origin we performed simulations of
Trapezium-equivalent stellar clusters (2500 stars in a virialized
$0.25$\,pc to $1.0$\,pc radius) with various compositions of JMOs and
stars.

Models in which planets form in wide hierarchical circumstellar
orbits (model ${\cal SPP}$), as proposed by
\cite{2023arXiv231006016W}, produce many single free-floating planets,
but insufficient numbers of pairs. The ratio of single to pairs of
planet-mass objects in these models are too low by a factor of 50 to
400, irrespective of the initial stellar distribution function.

The models in which pairs of planetary-mass objects orbit stars in the
form of a planet-moon system (or binary planets, model ${\cal SPM}$),
produce a sufficent number of free-floating planetary pairs, and cover
the proper range of orbits.  In particular the models that start with
fractal initial conditions tend to produce a sufficent fraction of
\jumbos\, among free-floating objects ${\cal O}(0.1)$, which is close
to the observed value of $0.078\pm0.012$. In the Plummer distribution,
the number of stars that survive with at least one planet-mass objects
is considerable.  These cold Jupiters have a typical orbital
separation of $\langle a\rangle = 382\pm75$\,au, and rather high
$\langle e \rangle = 0.74\pm 0.08$ eccentricity.  For the \jumbos\,
these models generally predict low-eccentricities ($e\aplt 0.4$),
whereas others lead to thermalized distributions ($\langle e \rangle
\sim 0.6$).

For model ${\cal SPM}$ to produce a sufficient number of \jumbos\, it
requires planet-moon pairs to form in $\apgt 900$\,au orbits around
their parent star. Such wide orbits are exotic since the circumstellar
disks observed in the Trapezium cluster tend to be smaller than
400\,au. We therefore do not see how such wide planet-moon pairs can
form around stars. If, however, a population of cold $a\apgt 100$\,au
JMOs are found in the Trapezium cluster, we do consider this model a
serious candidate for producing \jumbos.  Investigating some of the
observed \jumbos\ in the images published in
\cite{2023arXiv231001231P}, we get the impression that some \jumbos\,
may have nearby stars, but a thorough statistical study to confirm
this correlation is necessary.  Model ${\cal SPM}$ can easily be
confirmed or ruled out by establishing the existence of those left
over (or dynamically formed) planetary systems.

Ruling out models ${\cal FFC}$, ${\cal SPP}$, and possibly ${\cal
  SPM}$, we are left with the simplest solution; \jumbos\, form
together with the single stars in the cluster.  This model reproduces
the observed rates and orbital characteristics (a bit by construction,
though); it can also be used to further constrain the initial
conditions of the cluster as well as the \jumbos.

We tend to prefer model ${\cal ISF}$ with a 0.5\,pc Plummer sphere
because it can be tuned rather easily to reproduce the observed
population of \jumbos\, and JMOs. On the other hand, we also consider
the equivalent fractal model (${\cal ISF}$\_Fr\_R050) a good candidate
so long as \jumbos\ were formed somewhat later in the clusters'
evolution, once the rate of violent encounters reduced.  If \jumbos\,
form within $\sim 0.1$ Myr of the stars, our results diverge with
observations, forming too few \jumbos\, and with too tight
orbits. Even so, one strong point of the fractal models is their
natural ability to prune wide \jumbos\, and their natural ability to
form Jupiter-Mass Triple Objects (JuMTOs).

Single free-floating planetary objects were discovered in abundance
before in the Upper Scorpius association (between 70 and 170
candidates) \cite{2022NatAs...6...89M}, but these were considered to
be single free-floaters.  With an age of about 11\,Myr
\cite{2022NatAs...6...89M}, Upper Scoprius is expected to be rich in
single Jupiter-mass free-floating planets, but binaries will be rare
as the majority will be ionized. We still could imagine that a few
\jumbos\, have survived until today.

Finally, we would like to comment briefly on the nature of the objects
observed. We wonder that, if these objects formed in situ, and
therefore not around a star, they would be deprived of a rocky core.
\jumbos\, and JMOs would then more resemble a star in terms of the
structure, rather than a planet. This may have interesting
consequences on their dynamics, their evolution, and when they
encounter another star (collisions in our simulations are relatively
frequent). In those terms, we also wonder to what degree the term
``planet'' is rectified at all, and maybe it is time to revive the IAU
discussion on the definition of a planet.

\section*{Software used for this study}

In this work we used the following packages: \texttt{python}
\cite{10.5555/1593511}, \texttt{AMUSE} \cite{2018araa.book.....P},
\texttt{numpy} \cite{Oliphant2006ANumPy}, \texttt{scipy}
\cite{2020SciPy-NMeth}, \texttt{matplotlib}
\cite{2007CSE.....9...90H}.

\section*{Energy Consumption}

For $8400$ hours of CPU time and an energy consumption rate of
12\,Watt/hr \cite{2020NatAs...4..819P} and an emission intensity of
0.283\,kWh/kg \cite{doi:10.1002/cpe.3489} for performing all
calculations in this manuscript, we emitted $\sim 357$\,kg of
CO2. This amounts to roughly 32.48 tree-years of Carbon
sequestration. Enough to charge 43\,426 smartphones.

\section*{Acknowledgments}
We are grateful to Veronica Saz Ulibarrena, Shuo Huang, Maite Wilhelm,
Brent Maas, and Samuel Pearson for being available to discussing
\jumbos\, with us.  We thank NOVA for support.


\begin{appendix}
  \section{Similarity between $r_{ij}$ and $a$}\label{Appendix:A}

  The comparison between simulations and observations is somewhat
  hindered by the different perspectives. Whereas dynamicists prefer
  to use Kepler orbital elements, from an observational perspective
  such data is not always available. In our current study, we try to
  compare populations of binaries with observed objects. The latter
  are projected separations, which do not directly translate in
  orbital elements without full knowledge of the 6-dimension phase
  space of the orbit. We therefore have to compare projected separation
  with what we prefer to use, the semi-major axis of a bound two body
  orbit.

\begin{figure}
    \centering
        \includegraphics[width=0.45\columnwidth]{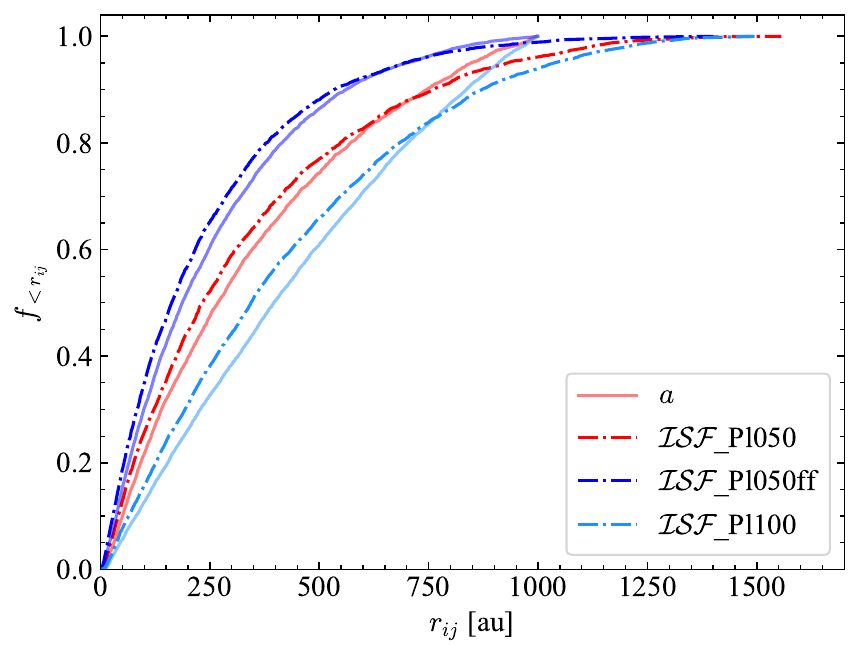}
        \includegraphics[width=0.45\columnwidth]{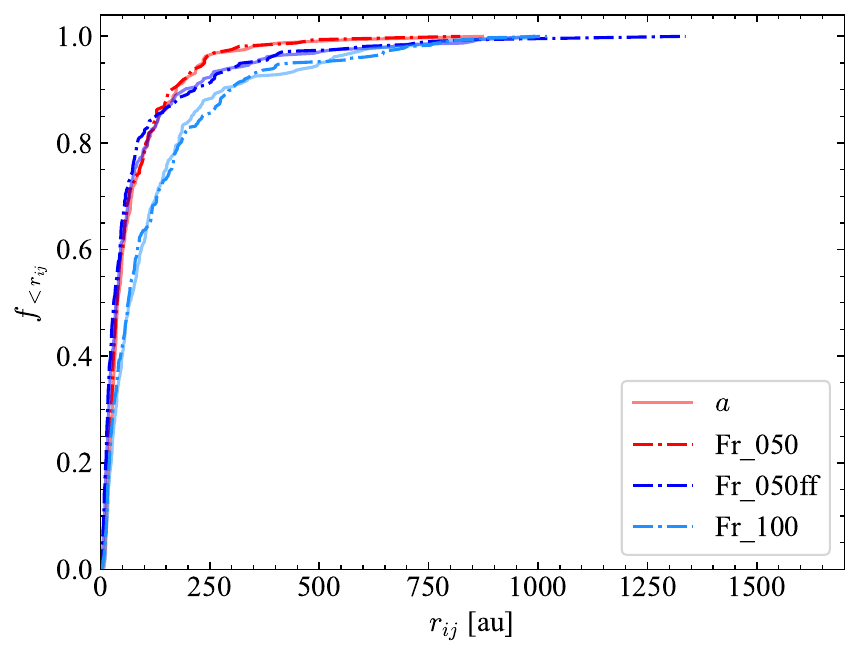}
        \caption{CDF of surviving \jumbo\, projected separation distribution for models Pl\_050, Pl\_050ff, Pl\_100 (left) and Fr\_050, Fr\_050ff, Fr\_100 (right). Overplotted are translucent lines denoting the respective models' \jumbos\, semi-major axis.}
         \label{Fig:Plummer_rsep}
   \end{figure}

  Both panels in figure \ref{Fig:Plummer_rsep} motivate
  our choice of analysing results in terms of the semi-major axis
  given the similarity between the curves.  In all cases, $r_{ij}$
  exhibits longer tails at shorter separations/orbits. However, these
  differences are so small, especially in the fractal case, and
  considering the width of the distribution (illustrated with the
  large values for the quartile intervals) we can safely interchange
  between one and the other. In doing so, we assume that the observed
  projected separation of \jumbos\, are equivalent to their semi-major
  axis, easing our discussion.
    
\end{appendix}
    
\end{document}